\documentclass{aa}  
\usepackage{graphicx}
\usepackage{multicol}
\usepackage{multirow}
\usepackage{xcolor}
\usepackage[draft,inline,nomargin,index]{fixme}
\usepackage{amsmath}
\usepackage{adjustbox}
\usepackage{rotating}
\usepackage{lscape}
\usepackage{longtable}
\usepackage{tabularx}

\begin{document} 
{

   \title{CHARA/SPICA: The six-telescope visible combiner and near-infrared fringe tracker for the CHARA Array}
    \titlerunning{CHARA/SPICA}
    \authorrunning{Mourard D. et al.}

   \author{D. Mourard \inst{1} \and
          P. Bério\inst{1} \and
          C. Bailet\inst{1} \and
          A. Caci\inst{1} \and
          J. Dejonghe\inst{1} \and
          P. Geneslay\inst{1,2} \and
          S. Lagarde\inst{1} \and
          D. Lecron\inst{1} \and
          A. Meilland\inst{1} \and
          F. Morand\inst{1} \and
          N. Nardetto\inst{1} \and
          C. Pannetier\inst{1,3} \and
          K. Perraut\inst{4} \and
          S. Rousseau\inst{1,5} \and
          D. Salabert\inst{1} \and
          N. Ebrahimkutty\inst{1} \and
          R.V. Iba\~{n}ez Bustos\inst{1} \and
          J. Jon\'{a}k\inst{1} \and
          R. Ligi\inst{1} \and
          H. Nowacki\inst{1} \and
          F. Patru\inst{1} \and
          M. Vrard\inst{1} \and
          L. Bourgés\inst{6} \and
          G. Mella\inst{6} \and
          N. Anugu\inst{7} \and
          T. ten Brummelaar\inst{7} \and
          C. Farrington\inst{7} \and
          D. Gies\inst{7} \and
          J. Jones\inst{7} \and
          R. Koehler\inst{7} \and
          K. Kubiak\inst{7} \and
          C. Lanthermann\inst{7} \and
          R. Ligon\inst{7} \and
          G. Schaefer\inst{7} \and
          N. Scott\inst{7} \and
          N. Turner\inst{7} \and
          J. Monnier\inst{8} \and
          S. Kraus\inst{9} \and
          J.B. Le Bouquin\inst{4} \and
          T. Gardner\inst{8,9} \and
          M. Gutierrez\inst{8} \and
          N. Ibrahim\inst{8} \and
          E. Aristidi\inst{1} \and
          Y. Caujolle\inst{1} \and
          C. Giordano\inst{1} \and
          A. Ziad\inst{1}
          }

   \institute{Universit\'e C\^ote d'Azur, Observatoire de la C\^ote d'Azur, CNRS, Laboratoire Lagrange, France\\
              \email{denis.mourard@oca.eu}
         \and
   OGS Technologies, France
         \and
   UNIDIA, Observatoire de Paris, Université PSL, CNRS, 92190 Meudon, France
         \and
   Univ. Grenoble Alpes, CNRS, IPAG, 38000 Grenoble, France
         \and
Université Bordeaux, CNRS, LP2I Bordeaux, UMR 5797, F-33170 Gradignan, France
         \and
Univ. Grenoble Alpes, CNRS, IRD, INRAE, Météo France, OSUG, 38000 Grenoble, France
    \and
The CHARA Array of Georgia State University, Mount Wilson Observatory, Mount Wilson, CA 91203, USA \and
University of Michigan, Ann Arbor, MI \and
University of Exeter, Department of Physics and Astronomy, Stocker Road, Exeter, EX4 4QL, United Kingdom
}

\date{Received ....; accepted ...}
\abstract
   {The suite called Stellar Parameters and Images with a Cophased Array (SPICA) has two interferometric instruments installed at the focus of the CHARA Array located at Mount Wilson, CA. SPICA is made of SPICA-VIS, a fiber-fed six-beam visible spectrograph with three spectral resolutions, and SPICA-FT, a six-beam near-infrared fringe tracker for the fast stabilization of the fringes.}
   {SPICA is opening access to imaging in the visible domain with an unprecedented angular resolution down to 0.2 milliarcseconds. It has been designed around a large survey of fundamental parameters of stars over the Hertzsprung-Russell diagram, aiming at understanding the deviations from the standard empirical relations of stellar physics as a function of activity: limb darkening, multiplicity, rotation, winds, and environments.}
   {SPICA makes use of the advanced technologies in electron multiplying detectors in the visible and electron-avalanche photodiode arrays in the near-infrared. It benefits from the newly commissioned adaptive optics on the one-meter telescopes of the array. The modules of the visible instrument, SPICA-VIS, optimize the injection of light into single-mode fibers for spatial filtering before spectral dispersion in the image plane. The fringe tracker, SPICA-FT, performs group-delay and phase-delay tracking for six beams in the H band. SPICA-FT can use an all-in-one or ABCD encoding of the fringe signals.}
   {We present the results of the SPICA commissioning and discuss the initial performance that has been established. In low-resolution mode, SPICA reaches a limiting magnitude of 6.5, consistent with the initial estimate without fringe-tracking stabilization.}
   {SPICA is operational on sky and is close to reaching the expected performance in low spectral resolution, in particular, for the Interferometric Survey of Stellar Parameters (ISSP). More work is still needed to achieve the ultimate performance in terms of sensitivity and to allow operations with higher spectral resolutions.}


\maketitle
\nolinenumbers

\section{Introduction}
\label{introduction}
After 50 years of development, long-baseline optical interferometry has reached a high level of maturity and offers a unique tool for probing the Universe to the broad astronomical community, from exoplanets to galaxies. The landscape today is dominated by the European Southern Observatory Very Large Telescope Interferometer (ESO/VLTI, \citealt{vlti}) in Chile and by the Center for High Angular Resolution Astronomy Array (CHARA, \citealt{chara}) in California. Magdalena Rigde Optical Interferometer (MROI, \citealt{mroi}) has recently obtained its first fringes and will ultimately offer a very powerful facility for imaging with ten telescopes over baselines of 340 meters. In the early ages of optical interferometry, access to sensitive low-noise large-format detectors has favored the initial steps in the visible. However, with the obvious interest of moving to larger apertures, the limitations generated by atmospheric turbulence, and the strong progression of the performance of infrared detectors, the domain has been dominated by instruments operating from the H to the N band. Consequently, about 90\% of the publications\footnote{https://www.jmmc.fr/bibdb/index.php} cover the infrared part of the spectrum.

Optical interferometry in the visible has grown through the precursors Interféromètre à 2 Télescopes (I2T, \citealt{labeyrie74}), Sydney University Stellar Interferometer (SUSI, \citealt{susi}), MarkIII \citep{mark3}, Grand Interféromètre à 2 Télescopes (GI2T, \citealt{gi2t}), Cambrigde Optical Aperture Synthesis Telescope (COAST \citealt{coast}), and Navy Precision Optical Interferometer (NPOI, \citealt{npoi}). More recently, the Visible spEctroGraph and polArimeter (CHARA/VEGA, \citealt{Mourard2009,Mourard2011}) and the Precision Astronomical Visible Observations (PAVO, \citealt{pavo}) brought important results through the improved angular resolution permitted by the combination of the longest available baselines and the short wavelengths of operation (see, e.g., \cite{abaur,bigot,ligi,phiper,pavo2,sbcr}). Precision and sensitivity were limited \citep{spie2012}, however, and new developments have been identified to make progress in addressing the most challenging science cases. It was necessary to adopt the principles of spatial filtering (see, e.g., \citealt{smfib}) to the case of visible wavelengths. Two major achievements were obtained with the development of the Visible Imaging System for Interferometric Observations at NPOI (VISION, \citealt{vision}) and the development and on-sky operation of the Fibered spectrally Resolved Interferometer – New Design at the CHARA Array (FRIEND, \citealt{friend}). Following these successes and by developing adaptive optics for the CHARA telescopes \citep{charaao1,charaao2,anugu2020}, the road was opened for the development of a new-generation visible combiner called Stellar Parameters and Images with a Cophased Array (SPICA) at the CHARA Array, as initially presented by \cite{2017JOSAA..34A..37M}.

The CHARA Array is equipped with several beam combiners, including two six-telescope infrared instruments. Michigan InfraRed Combiner-eXeter (MIRC-X, \citealt{mircx}) operates in the H band, while Michigan Young STar Imager at CHARA (MYSTIC, \citealt{mystic}) is designed for the K band. From the very beginning, the development of a new six-telescope visible combiner for the CHARA Array has been made in synergy for science and instrumental purposes with these two existing instruments. We also integrated the important progress obtained with the fringe tracker \citep{lacour} of the GRAVITY instrument at the VLTI \citep{gravity} into our design. From an astrophysical point of view, opening the possibility of simultaneous observations in the R, H, and K bands has clearly guided our choices.

The SPICA instrument, composed of the visible spectrograph SPICA-VIS and the near-infrared fringe tracker SPICA-FT, was commissioned in 2022 and 2023 and started its science life at the end of 2023 \citep{spica2024}. We provide a detailed overview of SPICA-VIS in its current state at the CHARA Array, and we succinctly describe SPICA-FT, which will be the subject of a separate paper. Section \ref{sec:science+specs} presents the main science drivers and the high-level specifications of the instrument. Section \ref{sec:instru} details the instrument concept and the different subsystems. Section \ref{sec:operation} describes SPICA operations, including the fringe-tracking capability. The data flow and the pipeline are presented in Section \ref{sec:data}. Section \ref{sec:perfs} describes the initial on-sky performance, and concluding remarks are given in Section~\ref{sec:conclusion}. Appendix\ref{sec:acronyms} summarizes the different acronyms used in this paper.

\section{Science drivers and specifications}
\label{sec:science+specs}
\subsection{Main science drivers}
\label{sec:science}
The main science drivers for visible interferometry were established a few years ago with a collaborative world-wide effort \citep{stee2017}. During the definition phase of SPICA, we decided to focus our attention on fundamental parameters of stars through an extensive and homogeneous survey of angular diameters. The detailed science cases of this Interferometric Survey of Stellar Parameters (ISSP) have been presented by \cite{spica2022}. 

The core program of SPICA is therefore related to the fundamental parameters of stars and planets. The combination of Gaia parallaxes \citep{gaia}, the high precision measurement of the transit of exoplanets with the PLATO space mission \citep{PLATO}, and the angular diameters measured with SPICA will provide the radius of exoplanets with 1\% precision. In addition, determining the radii of asteroseismology targets (using detailed modeling or scaling relations) will bring constraints on the mass and age of stars. Although asteroseismology alone is limited to a precision of $~10\%$ on these two main parameters \citep{lebreton2014}, combining asteroseismology and interferometry allows us to reach $~1\%$ \citep{lundkvist2025}. These measurements will be used to improve stellar evolution models and, ultimately, stellar ages. In addition, the imaging and spectroscopic capacities of SPICA will be used to study binaries (mass determination), stellar rotation, wind and environment, as well as limb-darkening. The aim is also to constrain the surface brightness color relations (SBCR) in different regions of the HR diagram and to study the impact of any stellar activity on the precision and accuracy of these relations. The SBCR is a powerful tool for deriving the angular diameter of any non-active star from magnitudes at two different spectral bands. Such relations are widely applied in various contexts, including the characterization of exoplanet host stars \citep{Gent2022, di_mauro_2022} and asteroseismology \citep{valle_2024, campante_2019}. They have also been used to determine distances to eclipsing binaries in the Large \citep{pietrzynski_2013, pietrzynski_2019} and Small \citep{graczyk_2020} Magellanic Clouds with unprecedented accuracies of 1\% and 2\%, respectively, results of particular importance for refining the Hubble–Lemaître constant \citep{riess_2022}.

Following the important space missions of the last decade in the domain of exoplanets through photometric measurements (KEPLER, TESS), it is remarkable that the upcoming missions (PLATO, ARIEL) are focusing more on brighter targets in order to achieve a very accurate determination of the stellar parameters. Thus, with the progress in sensitivity that is foreseen with the new visible combiner, the overlap of the samples is better in terms of the magnitude range, and the synergy between the science programs is very strong.

\begin{figure*}[ht]
  \centering
   \includegraphics[width=18cm]{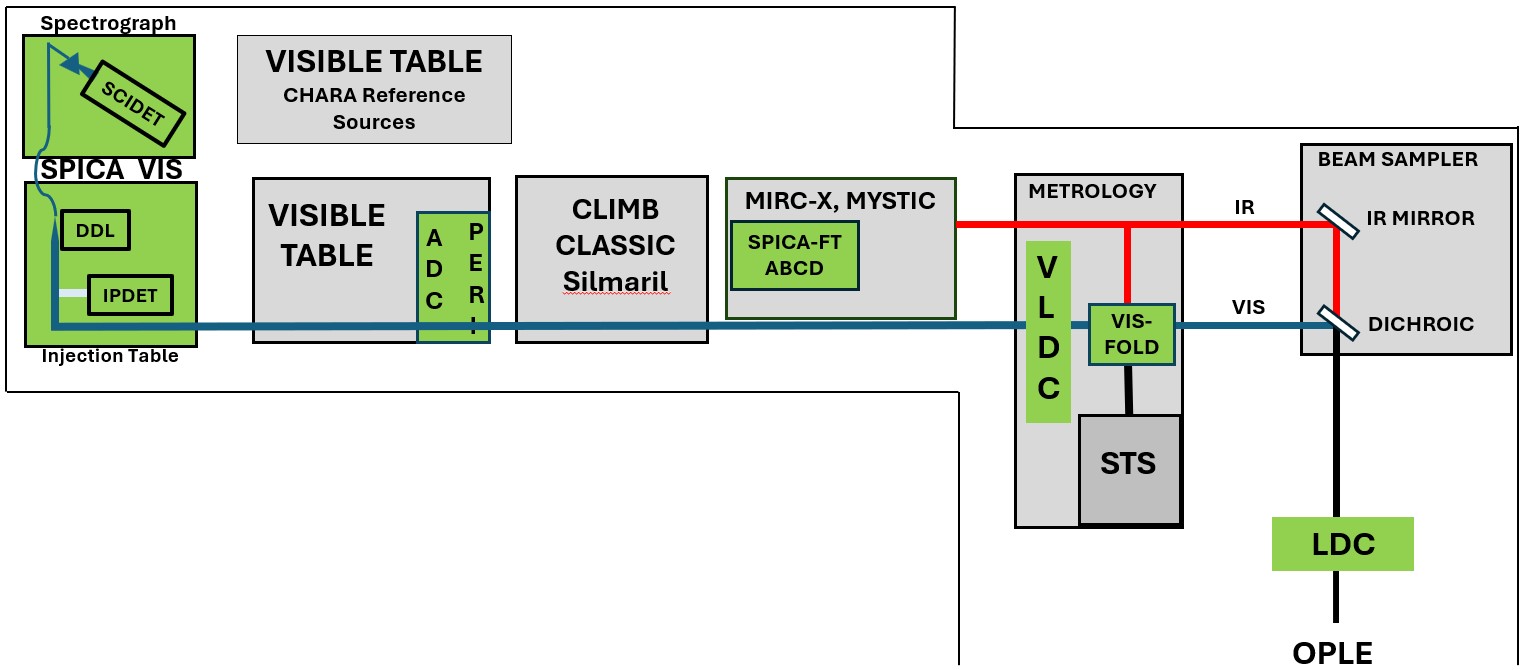}
   \caption{Different SPICA elements (in green) inside the CHARA beam-combiner laboratory. Visible light (in black and then blue) arrives from the telescopes and the delay lines (OPLE) in the bottom right corner. The infrared light (in black and then red) feeds the MIRC-X and MYSTIC instruments, also including the SPICA-FT ABCD chip. The periscope is used in down position to lift up the beams above the CHARA visible table and to feed the SPICA table. When the periscope is in up position, the beams do not propagate to SPICA, but the reference beams from CHARA visible table can propagate to CHARA telescopes for alignment purposes. The atmospheric refraction is corrected by the ADC. On the injection table, the light is injected in single-mode fibers transporting the beams to the spectrograph. The IPDET detector is used to align the image and pupil planes with a fraction of the visible flux, and the SCIDET detector records the dispersed fringes. The four elements LDC, VLDC, OPLE, and DLL are used to correct the longitudinal dispersion and cophase all visible and infrared instruments. The VIS-FOLD device injects light from the Six Telescope Simulator (STS) for alignment and cophasing purposes. }
   \label{fig:chara_spica}
\end{figure*}

\subsection{Instrumental specifications}
\label{sec:specs}
It is not surprising that the top-level specifications for this new instrument are related to sensitivity and spectroscopic capabilities. There is a long tradition in optical interferometry to pursue spectrally resolved, high-angular studies while, at the same time, increasing the coherence length of the instrument for easier operations. For SPICA, it was determined at an early stage that a low-resolution mode ($R\simeq 140$) and full coverage of the visible band were necessary to achieve the sensitivity required for the fundamental stellar parameters program. Scientific objectives concerning the study of the dynamical aspects of stars and their environment lead to the need for higher spectral resolving power. As a result of the compromise between scientific objectives and possible performance, our decision was to design a spectrograph capable of combining low resolution and two higher modes: $R\simeq 4300$ and $R\simeq 13300$.

With the progress of the CHARA adaptive optics program, spatial filtering with single-mode fibers is also critical to reach high precision on interferometric observables. As a consequence, dedicated effort was put into the injection of light in the single-mode fibers for efficient operation in the visible. Early simulations have shown that a gain in the injection level is possible if a real-time fast tip-tilt system is added in front of the fiber to correct for the residual motion due to the partial correction by the adaptive optics system. To cover the largest possible spectral band in the injection process, a correction of atmospheric refraction is necessary. Considering the interferometric signal, the differential birefringence of the fibers needs to be corrected, and the differential paths in air in the CHARA Array need to be corrected by longitudinal dispersion compensators, from the R band to the K band for an optimal use of all three beam combiners.  

The main conclusion of the preliminary studies for the estimations of the signal-to-noise ratio was that a fringe tracker that will allow long exposures in the visible is mandatory to achieve the performance for the science programs described in Sect.~\ref{sec:science}. In parallel to the development of SPICA-VIS, important efforts have been devoted to the development and operation of a six-telescope H band ABCD fringe sensor, and a complete optical-path-difference controller aimed at performing group delay and phase delay tracking at a frequency of about $250~Hz$ using the main CHARA delay lines.

\section{Instrumental concept}
\label{sec:instru}
 
 Two main systems make up SPICA: SPICA-VIS, the visible instrument, and SPICA-FT, a new fringe sensor installed in the MIRC-X instrument associated with a new optical path-difference controller for realizing group-delay and phase-delay tracking. SPICA is divided into different elements installed in the CHARA Beam Combiner Laboratory as presented in Fig.~\ref{fig:chara_spica}. The visible light (black and then blue) goes through the main longitudinal dispersion compensators (LDC), the visible longitudinal compensator (VLDC), the periscope and the atmospheric dispersion compensators (ADC). The SPICA-VIS injection table prepares the beams for injection into single-mode fibers through image-and pupil-plane adjustements made with an IXON897 ANDOR detector, called IPDET. Finally, the SPICA-VIS spectrograph table is used for the combination, dispersion, and recording (IXON888 ANDOR detector, called SCIDET) of the photometric and interferometric signals. The Six Telescope Simulator (STS) can feed SPICA through dedicated fold mirrors (VIS-FOLD). Finally, an integrated optic six-beam ABCD combiner is installed on the infrared MIRC-X table with its own fibers connected to the MIRC-X injection modules and feeding the MIRC-X spectrograph to form the SPICA-FT sensor.

\subsection{CHARA alignment}
\label{sec:alignment}
The alignment of the CHARA Array is based on a visible light source (laser or white-light source) retrofeeding the coudé optical trains. The combination of close and far targets allows control of the alignment of the reference beams. Light propagates up to the telescopes, and through retroreflectors placed at different places in the coudé train, CHARA can be aligned by controlling the orientation and position of the beams. Finally, the white-light source, after retroreflection, feeds the instruments using the VIS/IR dichroics and the beam samplers. The instruments are thus aligned on the CHARA reference beams, and, finally, the STS can be aligned on the instruments for further purposes. Because of its location in the lab and the use of the periscope, SPICA-VIS cannot be fed directly by the CHARA reference beams and is aligned with the STS. SPICA-VIS receives the telescope beams or the STS ones using the periscope. The group of the six bottom mirrors of the periscope is removable to allow propagation of the CHARA reference beams when needed.  

\subsection{SPICA-VIS injection table}
\label{sec:spicavis}
The SPICA-VIS injection table is presented in Fig.~\ref{fig:injection}. The CHARA beams, after reflection by the periscope, arrive in the upper right corner of the figure and are prepared to be injected into the single-mode fibers at the lower left corner with the injection device (INJ), mounted on a translation stage for the compensation of differential delay (DDL). From the entrance of the table, each beam crosses the shutter (SHU), and the polarization compensation device (PDC); it is then focused (FOC) to the image plane where a field lens (PUP) is installed for the control of the position of the beam. The field lens reimages the pupil plane on the fast tip-tilt mirror (FTT), and then the beam is collimated again (COL). Part of the light is directed by beam splitters (BSP) to the control camera, where a device (PIS) is used to image either the image plane or the pupil plane. When the fibers are back illuminated, part of the light is reflected by the backside of the BSP and sent back to the control camera by retroreflectors (RFL). During the commissioning of SPICA, the characterization of the BSP demonstrated that these plates behave correctly ($10\%$ in reflection, $90\%$ in transmission) across the spectral band $[600;900]$~nm. However, they also reflect full light below $600$~nm generating CONTROL images dominated by the short wavelengths, which degrades the fast tip-tilt performance (see Sect.~\ref{sec:pupim}). The addition of a high-pass filter ($>600$~nm) solved the issue.

\begin{figure}[ht]
  \centering
   \includegraphics[width=9cm]{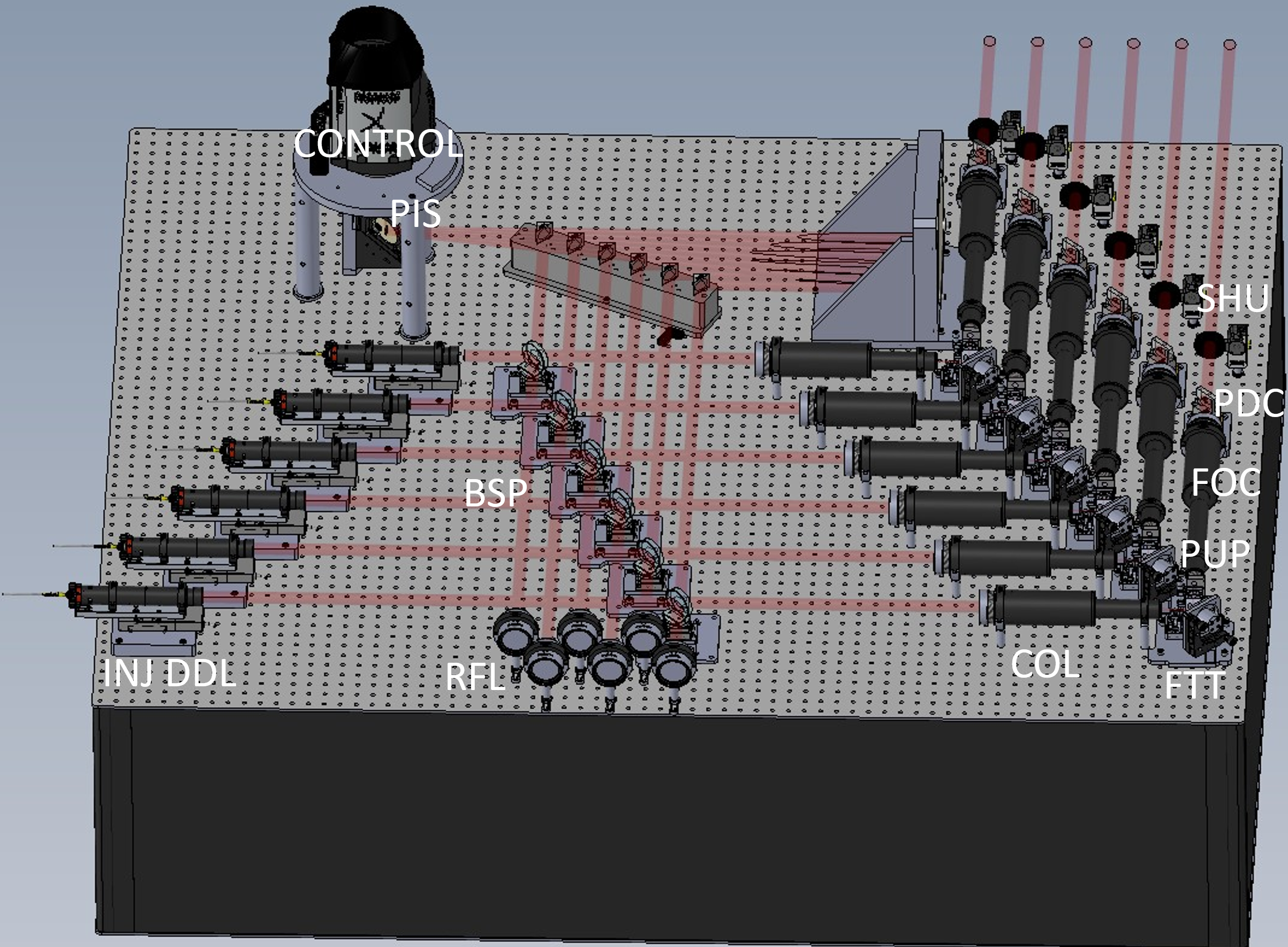}
   \caption{3D drawing of SPICA-VIS injection table presenting the main modules from the arrival of the CHARA beams (upper right corner) to the injection into the single-mode fibers (lower left corner). } 
   \label{fig:injection}
\end{figure}

\subsubsection{Compensation for longitudinal dispersion}
\label{sec:ldc}
Due to the propagation in air in the OPLE part of the CHARA coudé trains, the longitudinal dispersion should be compensated to permit a permanent cophasing of the instruments spread over the R, H, and K bands and also to optimize the contrast of the fringes within each band. This cophasing is made with the combination of M-LDC compensating the dispersion in the band of the fringe tracker (H band), differential delay lines for the K-band, additional visible compensators and differential delay lines in the visible. This complex scheme has been studied in detail by \citet{ldc} and is used during SPICA operations.

The VLDC are also used to compensate the residual chromatism of the SPICA instrument. It was determined that with all VLDC set to the same thickness, some of the fringe systems exhibited a residual dispersion on the STS fringes. It is generated by the difference in thickness of the PDC glasses (see Sect.~\ref{sec:pdc}) due to their different orientations, as well as some differences in the fiber lengths. The maximum correction applied is a difference of $450~\mu$m in thickness between two VLDC.\\

This longitudinal dispersion compensation scheme replaced the previous installation \citep{berger2003} at the end of 2023. The correct cophasing between the H band and the R band was demonstrated and the corresponding constant offsets were determined. However, we interpreted the low contrast performance of SPICA as caused by problems with the delay lines (see Sect.~\ref{sec:snr} and \cite{Anugu2026}), and it was only after correcting this issue in 2025 that we identified a remaining issue in the longitudinal dispersion compensation. Although being able to do efficient phase-delay tracking with SPICA-FT (see Sect.~\ref{sec:spicaft}), we found that the H band group delay was drifting, clearly indicating a residual dispersion. Both aspects (low contrast and group delay drift) were correlated and related to wrong numerical entries for the refraction index of air. The LDC and VLDC corrections were, in fact, adding dispersion instead of correcting it. This problem was solved in early 2026, and the recent results demonstrate a much better performance, as we explain in Sect.~\ref{sec:perfs}. 

\subsubsection{Refraction compensation}
The design goal for SPICA-VIS was to cover the red part of the visible band $[600;900]$~nm. It can be calculated that for a zenith angle of $45^{\circ}$, the refraction over this visible band is on the order of $0.37''$, which is much larger than the field of view of single-mode fibers in the visible. In order to correctly couple the whole spectral band in the fibers, the refraction needs to be compensated by the atmospheric dispersion compensators (ADC). Each ADC is made of two adjacent prisms, each with an angle of $1^{\circ}$. Both can rotate in opposite directions to generate a prism of variable angle between $0^{\circ}$ and $2^{\circ}$. The direction of dispersion is adjusted to match the field rotation of the telescope by rotating the two prisms in the same direction. Changing the angle of the ADC generates a deviation of the beam which is compensated by a tip-tilt of the upper mirror of the periscope (called IMG). These IMG mirrors are motorized with precision close-loop piezo actuators that permit blind correction of the deviation, taking into account, after calibration, the possible runout of the motion of the prisms. Their on-sky commissioning permitted us to calibrate the orientations and the field-rotation laws. We observed Sirius at a zenith angle of $51^{\circ}$, corresponding to a differential atmospheric refraction of $0.46''$. This effect was corrected using the ADC set to an angle of $0.94^{\circ}$, combined with a field rotation adjustment (for the hour angle of observation) of $-97^{\circ}$. A comparison of the injected flux before and after correction shows an improvement by a factor greater than $2$. The Zemax model of the injection device, accounting for this atmospheric refraction, indicates that when image stabilization is performed at 700~nm, the injection efficiency at 750~nm is reduced by a factor of three, while it becomes nearly zero below 650~nm and above 750~nm. The routine use of the ADC is therefore planned.

\subsubsection{Compensation for polarization}
\label{sec:pdc}
\begin{figure}[ht]
  \centering
   \includegraphics[width=9cm]{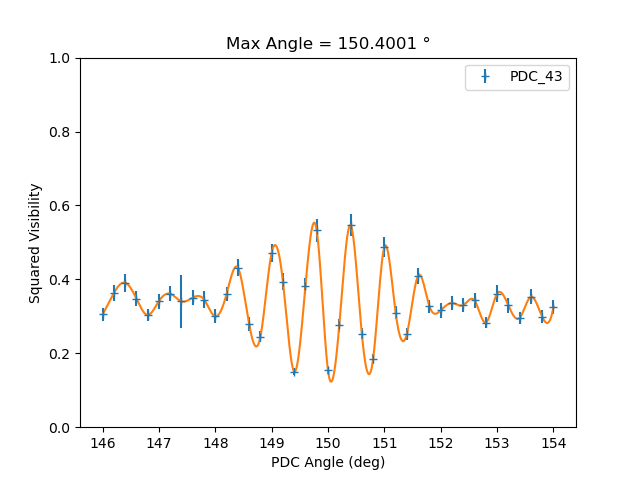}
   \caption{Squared visibility measurements on STS (blue crosses) as a function of the position angle of PDC3, while PDC4 is fixed. The maximum of contrast, determined by a numerical model (orange curve) of the measurements, gives the optimal position that compensates for the differential delay in polarization.} 
   \label{fig:pdc}
\end{figure}

Because of the differential birefringence of the single-mode fibers used in SPICA-VIS, we installed Polarization Device Compensators (PDC) to equalize the differential delay between the two directions of polarization of any pair of beams. As described in \citet{friend}, we use $\alpha$-Barium Borate ($\alpha$BBo) 5~mm-thick plates that can be precisely rotated to change differential polarization. One is fixed (on Beam 4), and the five others are adjusted to optimize the contrast in natural light. An example of the calibration process is presented in Fig.~\ref{fig:pdc} with a numerical model superimposed on the measurements. Small changes in position are needed from time to time, of the order of one tenth of a modulation. This is understood as a result of variations in the fibers or small changes in the STS mirrors. Positional variations are correlated with changes in temperature, especially after intensive laboratory work. Because of non-common paths between STS and SKY, it is necessary to check the possible changes in the PDC settings for SKY operation. After sky tests on October 1, 2025, we found that PDC3 and PDC6 needed small adjustments to maximize contrast. These settings are now part of the standard operations at each night.

\begin{figure}[ht]
   \center
\includegraphics[width=8cm]{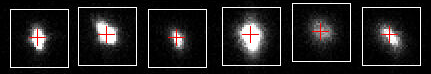}\\
\includegraphics[width=8cm]{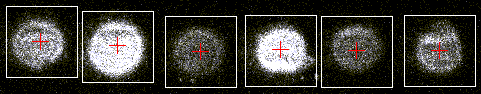}
   \caption{Top: Image planes of the six telescopes as seen on the IPDET detector (B1 to B6 from left to right). The differences in shape for the six beams are the results of static aberrations in the STS beams. Bottom: Same for the pupil planes. The images are from the STS source.}
   \label{fig:ipdet}
\end{figure}

\subsubsection{Image and pupil control}
\label{sec:pupim}
Aligning the CHARA beams entering in the injection table (see Fig.~\ref{fig:injection}) to correctly form the image of the star at the entrance of the single-mode fiber is one of the main critical aspects of the coupling between the instrument and the infrastructure. It has to be done before the beginning of each observing night, and adjusted after each slew to a new star. For doing that, we have implemented devices permitting the adjustment of the image plane (tip-tilt of the beams) and the pupil plane (translation of the beams), as well as the reimaging of the pupils within the SPICA injection table. The upper mirrors of the periscope (IMG) are motorized and permit a rough alignment in the image plane. However, as these mirrors are not located in a pupil plane, changing their orientations generates small translations of the pupil, of the order of one tenth of the pupil diameter for a correction of one Airy disk. To correctly align the CHARA pupil plane at the entrance of the injection module, we form an image plane using a focusing device (FOC), and a field-lens (PUP) is placed in this image plane on an XYZ translation stage. The Z-axis reimages the pupil plane on the next mirror (FTT) mounted on a fast tip-tilt actuator. The XY-axes correctly center the beam at the entrance of the injection module. After the FTT, a lens (COL) collimates the beam. The correct procedure is thus to align the image plane and then the pupil plane.

\begin{figure*}[ht]
   \center
\includegraphics[width=6cm]{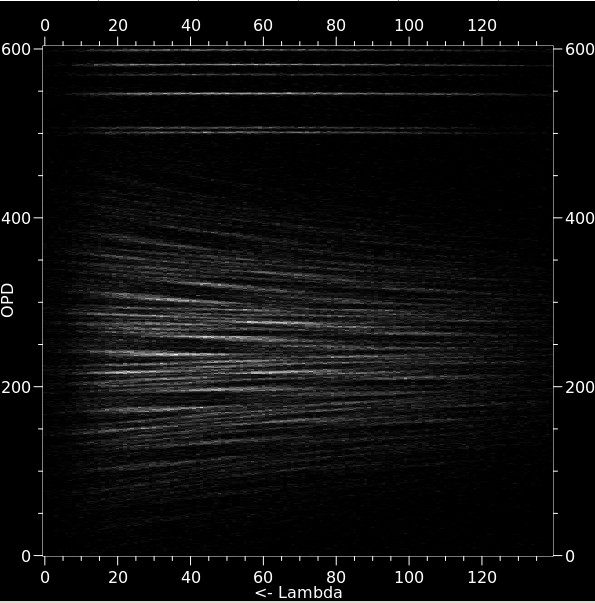}
\includegraphics[width=6cm]{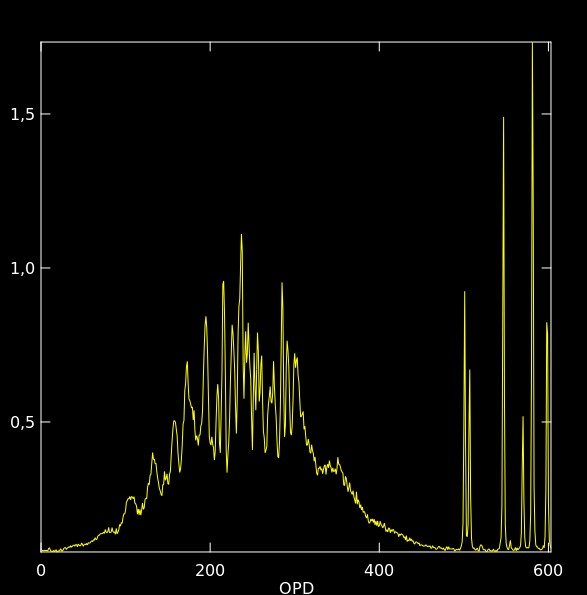}
\includegraphics[width=6cm]{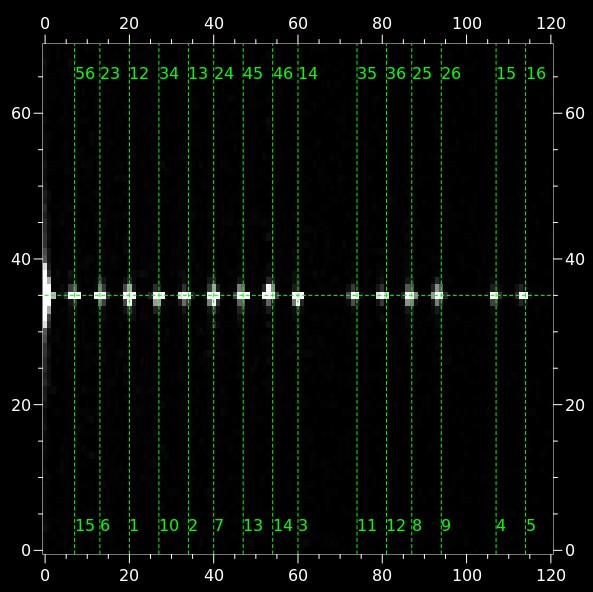}
   \caption{From left to right: 2D image of the SCIDET SPICA-VIS detector, 1D projection, and 2D power spectrum. The left panel shows the dispersed fringes, superimposed in the interferometric channels, and the 6 dispersed photometric channels (B1 to B6, from top to bottom) can be seen. In the middle figure, the 1D integration over the columns shows the non-redundant spacing of the six photometric channels (like the six fibers in the V-groove, B1 to B6 from right to left) and the Gaussian distribution of the interferometric channels, where the six beams are superimposed and interfere when the optical path is adjusted. The right figure presents the 2D power spectrum of the left image (with a rotation of $90^\circ$); the 15 fringe patterns are correctly phased when the energy is centered on the horizontal line, and the 15 fringe patterns are separated in frequency (horizontal axis) because of the non-redundant arrangement of the six fibers.}
   \label{fig:scidet}
\end{figure*}

\subsubsection{Injection and differential delay lines}
Once the beams are correctly aligned and prepared in terms of dispersion and polarization, injection into the single-mode fibers is the last operation of the injection table. The injection module is an optical tube that contains two lenses for the best focus over the spectral band. The head of the single-mode fiber is fixed on a Z-translation stage, allowing for a fine-tuning of the focus at the injection level. The whole optical system is mounted on a translation stage called the SPICA differential delay line (DDL). 
The injection modules were aligned, tested, and characterized before commissioning, and we found that they reached an injection efficiency of $~70\%$. The measurements with STS showed an injection level of $~40\%$, compatible with the initial setting if one considers the quality of the STS wavefront at the entrance of the injection module ($60$~nm RMS) corresponding to a Strehl ratio of $~70\%$. The same situation applies for the beams coming from the telescope, due to the residual static aberrations not seen or not totally corrected by the adaptive optics.

\subsubsection{Control of alignment}
To control the position of the pupil and the image, as explained in Sect.~\ref{sec:pupim}, 10\% of the light collimated by the COL is directed by a beam splitter (BSP) toward an ANDOR Ixon897 detector, called the control camera of SPICA-VIS, or IPDET. With a series of small oriented prisms, the images of each of the 6 beams are arranged linearly on the detector as shown in Fig.~\ref{fig:ipdet}. A movable lens, called the Pupil Image System (PIS), switches to pupil imaging when the PIS is in the beam. With these devices, the alignment in the image and pupil planes is possible, and the FTT closed-loop could be run for the image stabilization at the entrance of the fibers. When needed, shutters (SHU) can be used to close any of the 6 entrance beams. \\

The positions of the image and pupil on the detector are calibrated using the fiber-back illumination device (FBI) located near the exit of the fibers on the spectrograph table. A laser feeds the fibers back to the injection table. While most of the light continues toward the CHARA beams, the backside of the BSP redirects 10\% of the light to the control detector using a retroreflector (RFL). Reference positions for images and pupils are recorded when the FBI is IN. Aligning the image and pupil planes of the telescopes or of the STS on these reference positions ensures the direct injection of light into the fibers. On sky, the residual aberrations introduced by the partial correction of the adaptive optics create images that are much larger than the core of the single-mode fiber. It is therefore not mandatory to refine the centering, and we only use fiber maps on the STS to adequately calibrate the position of the fiber cores.

\subsection{SPICA-VIS spectrograph}

\begin{figure}
  \centering
   \includegraphics[width=9cm]{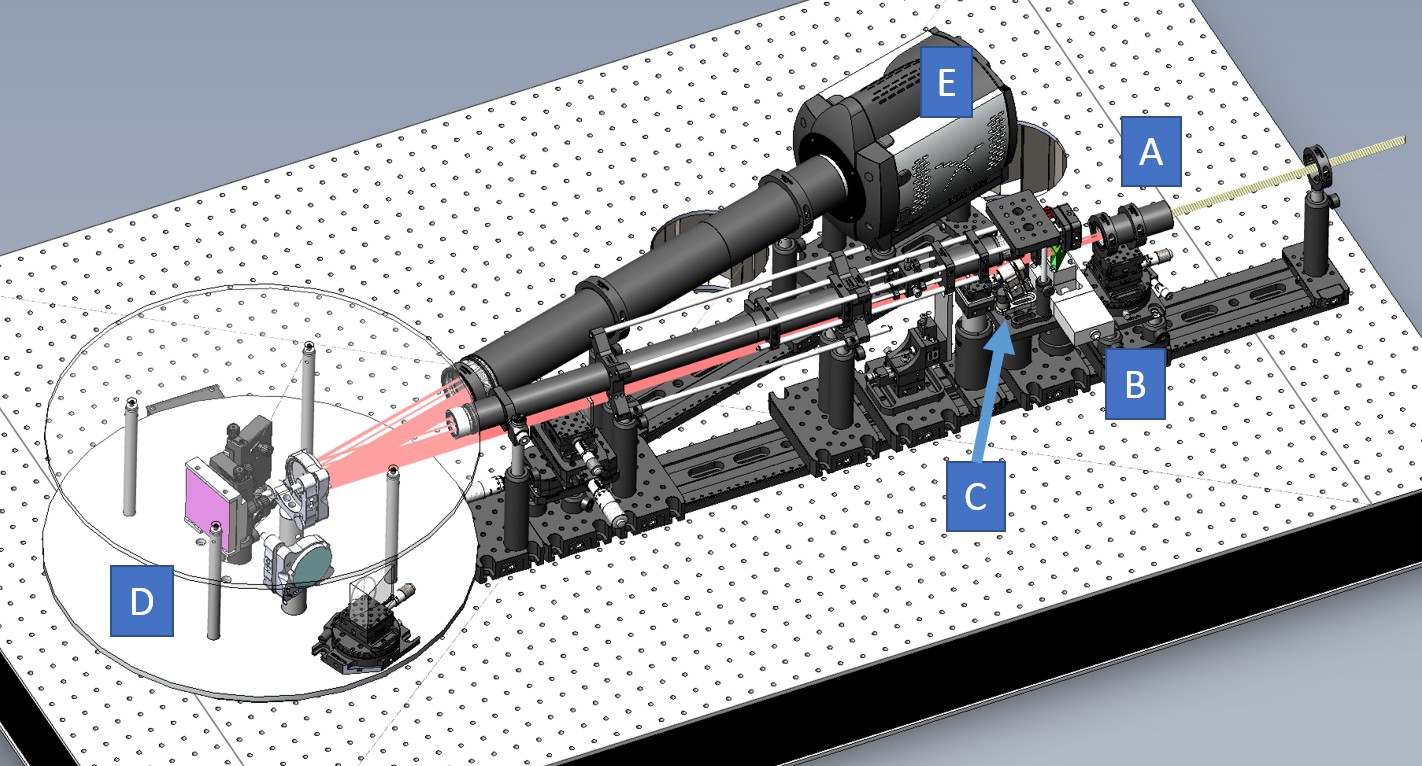}
   \caption{Drawing of the SPICA-VIS spectrograph. (A) is the V-groove with the different fibers arriving from the right of the figure, (B) is the FBI movable device, (C) is the beam splitter separating the photometric channels (going through the small tube in the direction of the dispersion (D)) and the interferometric channel (travelling in red, below the photometric tube), (D) is the dispersion device, and (E) is the ANDOR Ixon888 detector, equipped with a long tube containing the collimating lens.} 
   \label{fig:spectro}
\end{figure}

The SPICA-VIS spectrograph creates the science data on the IXON888 detector (see Fig.~\ref{fig:scidet}), called SCIDET. The entrance device of the spectrograph (see Fig.~\ref{fig:spectro}) is the V-groove for the positioning of the six single-mode fibers and the additional calibration fibers. The V-groove system was discussed previously by \citet{spica2022}. The six main single-mode fibers are arranged in a linear non-redundant scheme to permit a frequency separation on the detector. A microlens array is glued at the output of the V-groove to collimate the beams. About 10\% of the light is directed toward the photometric channels, whereas the rest of the light continues its travel to the interferometric channel. The photometric beams cross a series of lenses, inverting the pupil and the image planes before dispersion. In the interferometric channel, a pair of cylindrical lenses adapts the beam for a correct sampling of the fringes on the detector. The interferometric channel and the six photometric channels are then dispersed and collimated by a F-500mm lens on the SCIDET detector. The interferometric channel is collimated in the image plane (superposition of the Gaussian distribution of the six fibers), whereas the six photometric channels are separated on the detector, due to the conjugation with the microlens array plane at the exit of the V-groove. \\
For higher dispersion, we have two different gratings with a resolving power of R=4300 and 13300, over 500 spectral channels on the full width of the detector. The medium resolution covers a band of $~88$~nm and five settings have been defined around a central wavelength of $640, 685, 720, 800,$ and $850$~nm. The high-resolution spectral band is $~29$~nm wide and the central wavelength could be specified at any value within the instrument range. The low resolving power mode (R=140) is provided by the dispersion of a prism after reflection on a flat mirror. The whole $[600,900]$~nm band is covered in 60 spectral channels spanning 120 columns of the science detector. This mode is the most sensitive one and has been used almost exclusively for the different science programs.

\subsection{SPICA-FT}
\label{sec:spicaft}
The SPICA-FT has been described in detail by \citet{spicaft}. The FT is composed of three main elements: 1) the fringe sensor measuring the phase of the 15 fringe systems, 2) a calculator computing the commands to be sent to stabilize the fringes, and 3) the actuators for the correction of the optical path differences, which are the main CHARA delay lines. CHARA implemented some time ago a fast and dedicated communication channel to the delay lines controllers for the fringe tracking \citep{Anugu2026}. We have developed a new calculator called $mirc\_opdcontroller$, as a new server embedded in the MIRC-X system. It reads, in a shared memory, the 6 fluxes and the 15 complex coherent fluxes extracted in real time from the frames acquired by the MIRC-X C-Red\,One detector. Following the design of the Gravity Fringe Tracker \citep{lacour}, we implemented an extension of the equations for 6 beams, 15 fringe patterns and 20 closure phases. The calculator computes in real-time the group delay, phase delay, signal-to-noise ratio, and variances. It manages all the state machine commands necessary to feed and control the combination of group delay and phase delay tracking, as well as the necessary relocking functions when the fringes are lost. The closuse phases are used to build phase reference vectors allowing to lock the fringes even when the star exhibits intrinsic phase signals on some baselines.\\

\begin{figure}[ht]
  \centering
   \includegraphics[width=7cm]{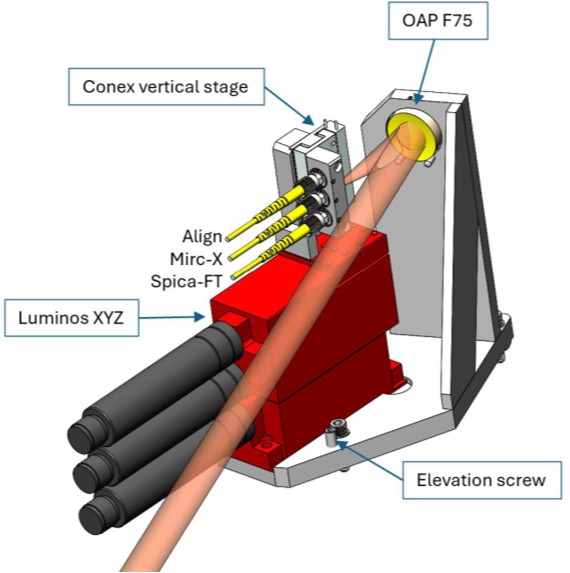}
   \caption{Design of the injection module of MIRC-X with the new translation stage that permits injection of the beam into different fibers, mainly all-in-one or ABCD combiners.} 
   \label{fig:injectionspicaft}
\end{figure}

The SPICA-FT is operational and can be fed by the classic All-In-One MIRC-X combiner or by the SPICA-FT integrated optics chip encoding each of the 15 fringes following the ABCD principle \citep{spicaft}. Switching from one combiner to another takes less than five minutes with realignment at preset positions. We have installed translation stages (see Fig.~\ref{fig:injectionspicaft}) in the focus of each off-axis parabola, allowing injection into different sets of fibers, one for All-In-One, one for ABCD, and one for alignment purposes. Presets of the XYZ stages for each of the 6 positioning systems at the focus of the off-axis parabola have been calibrated for the two recombination modes. Very few alignment steps are necessary when making the switch.\\

Running SPICA-FT is an operation on top of the classical MIRC-X operation. Fringes are first acquired with the classical MIRC-X group delay control. When fringes are centered, this first loop is open and the SPICA-FT group- and phase-delay tracking is activated. For a correct determination of the signal to noise thresholds, an estimation of the phase variances is done, a few times per night, by setting the delay lines out of coherence. Recording science data with MIRC-X could be done in parallel of the fast group- and phase-delay loops. 

Despite limitations due to issues in the CHARA delay lines control up to the end of 2024 as well as the wrong correction of longitudinal dispersion, SPICA-FT operations demonstrated stable phase tracking between $100$ and $180$~nm as presented in Fig.~\ref{fig:phasetracking}. 

The SPICA-FT is used routinely during ISSP runs and is typically working correctly up to magnitude H=6.5. During fast seeing episodes, the performance is degraded so that the phase-tracking mode is failing. SPICA-FT can be run up to a frequency of 500~Hz, but it has been discovered that the performance degrades. Analyzing the residual and the highest possible gain of the phase-tracking loop, as a function of the sampling frequency, we identified a loop-delay larger than 5~ms at 300~Hz and 500~Hz. The most critical remaining action on SPICA-FT is thus to understand this loop-delay and reduce it to improve the overall performance. In addition to this work, the remaining SPICA-FT activities include: (1) the development of a more user-friendly graphical interface for CHARA observers; (2) the characterization of the overall performance, in terms of magnitude and phase residuals, for both the All-In-One and ABCD combiners, together with the acquisition of seeing parameters (see Sect.~\ref{sec:psaum}); (3) the validation of tracking performance when the SPICA-FT loops are fed by MYSTIC; and (4) the delivery of a post-processing pipeline for MIRC-X data acquired with the ABCD combiner.

\begin{figure}[ht]
  \centering
   \includegraphics[width=9cm]{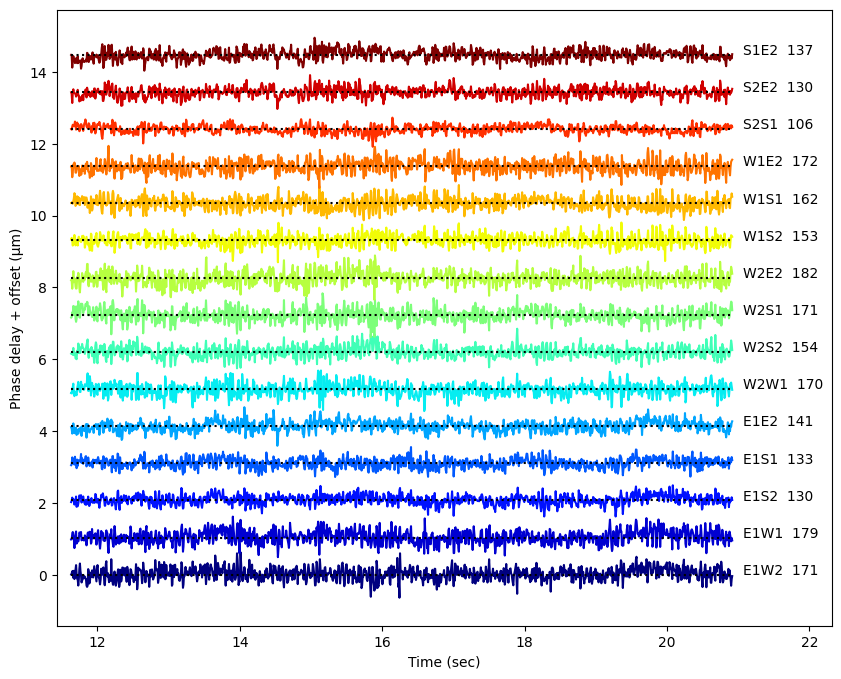}
   \caption{Screenshot of the SPICA-FT real-time display showing the phase of the 15 baselines as a function of time. The star we observed on 2025 July 9 was HD~3360. The values on the right gives the RMS of the phase residuals in nanometers. The residuals range between $100$ and $180$~nm, corresponding to a stabilisation better than $\lambda/10$ in the H band, and about $\lambda/5$ at 700~nm. The figure represents about $10$~s of phase tracking. } 
   \label{fig:phasetracking}
\end{figure}

\subsection{PSAUM: A seeing monitor for CHARA}
\label{sec:psaum}
Adaptive optics and fringe tracking performance are critical elements in the optimization of an interferometric array. For SPICA operating in the visible at a mean wavelength of $700~nm$, it is of utmost importance to reach the best behavior of these systems. Their characterization is usually done through a complete modeling in which the seeing conditions are among the main parameters. And finally, for their operational optimization, having an independent estimation of the seeing conditions is highly valuable. The lack of a seeing monitor at Mount Wilson Observatory has probably been one important limitation in recent years. In 2025, we installed an independent seeing monitor called PSAUM\footnote{patent US-2023-0072720} (Polar inStrument for Atmospheric tUrbulence Monitoring). This 10~cm aperture telescope without motorized mount is pointed at Polaris and automatically captures an image of the star with a wide field of view detector. Every minute, a 20s sequence of short-exposure images (5 and 10~ms) is taken. The analysis is done in real time and the different seeing parameters are then broadcasted on the CHARA network. Mean values over 10~min are usually formed to smooth the individual data and obtain the general tendency during the night. PSAUM gives an estimate of seeing, r0, t0, isoplanetism and isopistonic angle, and index of scintillation. A representative night is presented in Fig.~\ref{fig:psaum}.

\begin{figure}[ht]
  \centering
   \includegraphics[width=9cm]{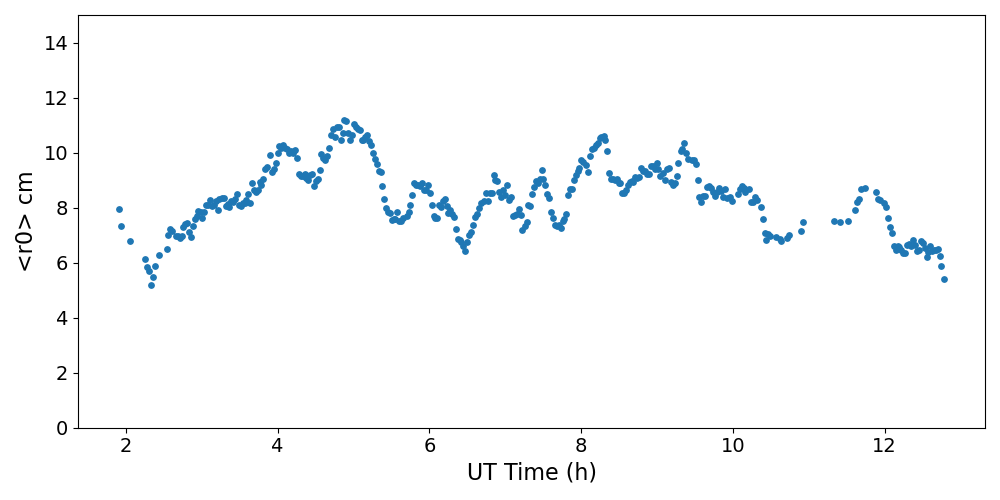}
   \caption{$10-~min$ average of r0 measurements by PSAUM at Mount Wilson Observatory for the night of 2025 December 15. The r0 values are given for a wavelength of $0.5\mu$m. } 
   \label{fig:psaum}
\end{figure}

\section{Operation}
\label{sec:operation}

\subsection{Preparing the observations}
Carrying out observations of a large stellar sample is not common practice in interferometry. The ISSP survey described in Sec.~\ref{sec:science} plans to observe 1000 stars (800 for angular diameters and 200 for imaging) over the course of several years. To make the best use of observing time, we have developed a tool called the Night Scheduling Software (NSS\footnote{https://github.com/dsalabert/spica-nss}). Its purpose is to choose, from a pre-selected pool of about 3000 stars, the targets to be observed in a given range of declination and magnitude, following a set of priorities based on the completion status of each subprogram. This strategy has two main advantages. First, it increases efficiency by favoring short slews, stable alignments, and consistent fringe offsets. Second, it ensures a stable and repeatable array configuration throughout the night, which in turn improves the stability of the instruments’ transfer function. The NSS tool is connected to the JSDC catalog\footnote{JSDC available at http://www.jmmc.fr/jsdc} for identifying calibrators; it also uses a sample of robust primary calibrators that are scheduled for observation from one night to another. The NSS tool is interoperable with ASPRO2, for the final preparation of the night strategy and its management. Finally, ASPRO2 sends the required information to CHARA and SPICA for data recording, using the interoperability principles of the A2P2 software.

The 3000 preselected stars correspond to an initial catalog called SPICA-DataBase (SDB) that is used for the follow-up of the progress of the survey. This survey management is done through a web-service\footnote{available at https://spicaweb.oca.eu}, connected to SDB and to an additional catalog called SPICA-FollowUp (SFU) storing the information about the observations carried out. We can add new targets into SDB, update the priorities, and validate or reject observation blocks. Additional python tools connected to the SDB and SFU catalogs perform statistical and additional global analyses with cross-matches with online catalog such as \textit{Gaia}.

\subsection{General software architecture}
The SPICA-VIS and SPICA-FT instruments are controlled using the CHARA client-server architecture. For SPICA-FT, we added, on the MIRC-X computer, a server (\textit{mircx\_opdcontroller\_server}) managing the group delay and phase delay algorithms. Two clients are used, one for the setting of the parameters of the loops (\textit{mircx\_opdc\_gtk}) and one for the real-time display of the different data (\textit{mircx\_opdc\_rtd}). These clients run on any remote machines in front of the user. A more user-friendly interface is currently being developed for general use by the CHARA community.\\
For SPICA-VIS, the server architecture is based on two computers, spica-control and spica-science. On the spica-science machine, we run \textit{spica\_scidet\_server} for the control of the science detector and for the acquisition of the science data, and \textit{spica\_ob\_server} is used to communicate the information of the next observing block between ASPRO2\footnote{Available at http://www.jmmc.fr/aspro} and \textit{spica\_scidet\_server} using the A2P2 interface\footnote{https://github.com/JMMC-OpenDev/a2p2}.\\
The spica-control machine is used for managing the Control Detector through a dedicated server (\textit{spica\_ipdet\_server}) and for the control of all the SPICA-VIS devices (Periscope, IMG mirrors, ADC, Shutters, PDC, PUP, FTT, DDL, PIS, FBI, Dispersion). Except for the FTT used in a close-loop system and the DDL adjusted for the chromatic correction in an open-loop system, all the other controls are just for alignment and nothing else is moving during the acquisition. A shell script is used at the beginning of the night to start or restart the servers. Moreover, we use two additional servers on the CHARA side for the control of the LDC and of the VLDC stages, and one additional server for calculating the air paths and the required glass thicknesses and offsets of the differential delay lines. 

Three main graphical user interfaces are used: \textit{spica\_ipdet\_gtk} for the control camera and the devices of the injection table, \textit{spica\_scidet\_gtk} for the science detector and the dispersion system, and finally, \textit{spica\_rtd\_gtk} for the real-time processing of the science detector frames to control the fiber injection and the appearance of the fringes.

\subsection{Standard steps for a complete SPICA observation}
\label{sec:observingsteps}
Before the start of the night, the SPICA-VIS instrument is aligned on the STS. A complete calibration sequence is thus recorded doing successively (see the different items of Sec.~\ref{sec:pipeline}) a background, a spectral calibration, a  $\kappa$-matrix calibration, and a STS fringes data set plus the foreground (out of coherence) with the associated darks. When the decision to slew to a new star is made, the following steps are taken, following the SPICA User's Manual\footnote{https://chara.gsu.edu/wiki/doku.php?id=chara:instruments}:
\begin{itemize}
    \item During the slew by CHARA, check the correct setting of the LDC, VLDC, CHARA delay lines and SPICA DDL. Set the ADC to their expected position.
    \item When the star is locked on the adaptive optics devices and aligned in the infrared on the Six Telescope Star Tracker (STST), center roughly the star in the corresponding box of the control camera by moving the required IMG mirrors. The STST is then used in guiding mode to stabilize the beam drift during the data recording. 
    \item Check and adjust if necessary the position of the pupil in SPICA control detector.
    \item Lock the fast tip-tilt mirror to center the star on the reference position and activate the low-frequency offload of the FTT to the IMG alignment mirrors. This step compensates the visible drift of the alignment with respect to the infrared alignment during the tracking and keeps the FTT close to the center of their active strokes. This drift is due to dispersive materials in the optical train of the CHARA telescopes. With the recent progresses brought by the slow infrared guiding in the laboratory (Six Telescope Star Tracker - STST), the need for offload is strongly reduced and it is activated only during one minute after the lock of the FTT loop. The impact on the pupil position is thus negligible if the initial alignment has been done correctly.
    \item Find the fringes with MIRC-X group delay tracker and close the group delay and phase delay loops with SPICA-FT. Co-phase MYSTIC with the associated differential delay lines.
    \item If necessary, adjust the co-phasing of the visible fringes in SPICA-VIS science detector using the DDL. Refresh the setting of the ADC.
    \item Start recording simultaneously SPICA, MIRC-X, MYSTIC data as well as the telemetry data of SPICA-FT (results of the real-time processing of each recorded frame) and a 30~s sequence of control images. For SPICA, the recording is made in ten files of 1 minute duration. MIRC-X (resp. MYSTIC) records 10 (resp. 9) files of 30~s (resp. 33~s) covering 5~min; then a remapping is done and a new sequence of 5~min is done.
    \item If required, finish with a shutter sequence on MIRC-X/MYSTIC for calibration purposes.
\end{itemize}

We do not describe the required calibration sequences with the STS for MIRC-X and MYSTIC here. They are described in the corresponding papers. When SPICA-FT is used with the ABCD chip, the real-time processing uses a pixel-visibility matrix calibrated on the STS. This calibration sequence records a dark, a shutter sequence, and modulated fringes for all pairs of beams. The same principle applies for the post-processing of the MIRC-X data when the ABCD chip is used.

\section{SPICA data flow and pipeline}
\label{sec:data}
\subsection{Data flow}

The SPICA data flow can be described in three main steps. The first concerns the processing of the raw data to obtain the uncalibrated L1 FITS files. The second is the calibration of the Science data to generate the calibrated L2 files. The third is the final validation of the L2 files for the programmatic aspects of the ISSP survey through the web-service, and their publication in the OiDB database\footnote{ISSP collection available at https://oidb.jmmc.fr/collection.html?id=issp}. 

During step 1, the raw SCIDET data are copied from the spica-science computer to the spica-pipeline machine, together with the SPICA-FT telemetries from the MIRC-X computer. The STS calibration files are processed to generate all the necessary calibration maps (dark, spectral dispersion law, kappa matrix, and fringe position). Details are given in Sect.~\ref{sec:pipeline}. After that, all individual 1~min science and calibrator raw data are processed and uncalibrated L1 files are created. An additional L1 file is created as result of the complete merging of the 10~min recording. A summary of the processing is stored in a quality control PDF file for each observing block. All L1 calibrator files are appended with the diameter information derived from the JSDC catalog \citep{jsdc2017}. Finally, the L1 and processing report files are copied to the SPICA data center in Nice and the spicaFollowUp catalog is updated with the new entries. The metadata (L0) are extracted by JMMC/ObsPortal tool from the header files and synchronized into the OiDB\footnote{L0 data at https://oidb.jmmc.fr/collection.html?id=issp\_l0}. Step 1 is not interactive and runs as a \textit{Python} script at the end of the observation night. A parallel archive of the raw data and of the L1 files plus processing reports is done toward the CHARA data center in Atlanta.

Step 2 (see Sect.~\ref{sec:calib}) is an interactive phase during which the PI of the data uses the tools of the SPICA pipeline to generate the L2 calibrated files. Different spectral or temporal binning is possible. The main program is the estimation of the transfer function as a function of time during the night, based on the L1 data for the calibrators. After that, the calibrated visibilities for the science targets are estimated and the L2 files are produced. All processing parameters are stored in the header of the L2 file, such as the calibrators used, the rejected baselines or spectral channels, and the version of the pipeline. This step is also applied on the MIRCX and MYSTIC data after their pre-processing at the CHARA data center in Atlanta and their transfer to the main center in Nice. The headers of these L1 data are updated to match the ISSP characteristics in terms of identifier in SPICA-DB, name of the PI, and reference of the sub-program of the ISSP survey. 

Step 3 is also interactive through a web-service. This tool allows us to follow the progress of a program by showing the existing L1 and L2 data (SPICA, MIRC-X, and MYSTIC). For the validation of a L2 file, the PI sets the "HIERARCH QUALITY LEVEL" keyword in the header and could add comments describing the processing and the calibration processes. The L2 file is shared on a web server with a secured access (PI or Co-I based upon the release date) and its metadata referencing the observation are synchronized with the OiDB SPICA collection. Finally, the completion rate of the corresponding program in the SPICA catalog is updated and the reference of the L2 file is stored. 

\subsection{SPICA-VIS pipeline}
\label{sec:pipeline}
The SPICA data reduction pipeline is implemented in {\it Python3.7} and produces science-ready data in OIFITS format \citep{oifits2}. The main steps of the pipeline are as follows.
\begin{enumerate}
\item The detector bias map is estimated from a cube of dark images. They are obtained by closing the shutter just before the detector.
\item The spectral calibration is done by feeding the spectrograph with two fibered spectral calibration sources (Neon and Mercury/Argon). The pipeline derives the pixel-wavelength relation in the topocentric frame by (i) identifying the well-known spectral lines of the calibration sources in the recorded spectrum and (ii) fitting a third-order polynomial function to the (pixel - wavelength) pairs. With this method, the precision of the wavelength calibration reaches 0.2\%. 
\item The interferometric signal is estimated in the Fourier space. It is therefore necessary to know the spatial frequency of the fifteen fringe peaks. For this, we record fringes with the STS artificial source. The pipeline computes the power spectrum of each spectral channel and estimates the spatial frequency of the center of the fifteen fringe peaks.
\item The $\kappa$-matrix gives the flux ratio between the interferometric and photometric channels for each beam. It is estimated by recording six image cubes corresponding to all shutters closed except one.
\item Squared visibilities for the baseline $ij$ are computed as follows ($<>$ represents the time average):
\begin{equation}
    V^2_{ij}(\lambda)=\frac{<|\tilde{I}_{ij}(\lambda,t)|^2>-\beta(\lambda)}{2\kappa_{i}(\lambda)\kappa_{j}(\lambda)<P_i(\lambda,t)P_j(\lambda,t)>},
\end{equation}
where $\tilde{I}_{ij}$ is the Fourier transform of the interferogram at the spatial frequency corresponding to the center of the fringe peak $ij$, $P_i$ is the photometry of the beam $i$, $\kappa_{i}$ is the coefficient of the $\kappa$-matrix for the beam $i$ and $\beta$ is the photon bias. This bias is measured at high spatial frequencies, where no astrophysical or instrumental signals can be present.
\item The closure phase of the triplet $ijk$ is computed as the argument of the bi-spectrum,
\begin{equation}
    C_{ijk}=<\tilde{I}_{ij}(\lambda,t)\tilde{I}_{jk}(\lambda,t)\tilde{I}_{ik}^*(\lambda,t)>-\beta_{ijk}(\lambda),
\end{equation}
where $\beta_{ijk}$ is a real additive term, accounting for bias and visibility and due to the all-in-one combination of the interferograms. Its estimation follows the method described in \cite{mircx}.
\item The differential phase $\phi_{ij}^k$ is defined as the phase of the fringes corresponding to the baseline $ij$ and of the spectral channel $k$, assuming the following constraint:
\begin{equation}
\sum_{k=0}^{n} \phi_{ij}^k=0,
\end{equation}
where $n$ is the number of spectral channels. The estimation is done in two steps: (i) computing the argument $\psi_{ij}^{k,k+1}$ of the cross-spectrum between two consecutive channels $<\tilde{I}_{ij}(\lambda_k,t)\tilde{I}_{ij}^*(\lambda_{k+1},t)>$, and (ii) solving the following linear system:
\begin{equation}
\begin{pmatrix}
1 & -1 & 0  & \cdots & 0\\
0 & 1  & -1 & \cdots & 0\\
\vdots & \vdots & \vdots & \vdots & \vdots \\
0 & 0  &  0 & \cdots & -1\\
1 & 1  &  1 & \cdots & 1\\
\end{pmatrix}
\begin{pmatrix}
\phi_{ij}^0 \\
\phi_{ij}^1 \\
\vdots \\
\phi_{ij}^{n-1} \\
\phi_{ij}^n \\
\end{pmatrix}
=
\begin{pmatrix}
\psi_{ij}^{0,1} \\
\psi_{ij}^{1,2} \\
\vdots \\
\psi_{ij}^{n-1,n} \\
0 \\
\end{pmatrix}.
\end{equation}

\end{enumerate}
For steps 5, 6, and 7, a selection of frames is applied before estimating the interferometric signals. This selection is based on the photometry of each telescope and on the detection SNR of the fringes given in the telemetry of the fringe tracker. For a visibility measurement on baseline $ij$, the frame is valid if both fluxes are above the photometric threshold and if the SNR of SPICA-FT for the corresponding frame is above the detection threshold. For a closure phase measurement on triplet $ijk$, the frame is valid if the three fluxes are above the photometric threshold and if the SNR of SPICA-FT for the corresponding frame and the three baselines $ij$, $jk$, and $ik$ are above the detection threshold.\\

\subsection{Calibration tool}
\label{sec:calib}
The last step of data reduction consists in calibrating the raw visibility, which means removing the instrumental signature. Actually, the raw contrast is always lower than expected from a theoretical point of view. The raw visibility $V_\mathrm{raw}$ could be modeled as follows:
\begin{equation}
    V_\mathrm{raw}=V_\mathrm{inst}V_\mathrm{exp},
\end{equation}
where $V_\mathrm{inst}$ is the system visibility (or transfer function) and $V_\mathrm{exp}$ is the expected visibility. The transfer function is estimated on unresolved stars or on stars with a well-known geometry, called calibrators. The transfer function is derived by dividing the raw visibility of the calibrators by their expected visibilities assuming a central uniform disk (UD) model. UD diameters are automatically retrieved from the JMMC catalog \citep{chelli2016,jsdc2017} or can be given manually. Then the transfer function estimates are interpolated at the time of the science observation following the method proposed in \cite{lachaume2019}.

This function has been adapted to MIRC-X and MYSTIC data. All processing parameters are stored in the headers of the L2 OIFITS file. For sensitivity purposes, spectral binning could be done before the calibration. 

Interferometric data are strongly subject to correlations within the data, due to the different spectral channels, the different baselines, and the fact that the same calibrator is used for correcting the transfer function. We implemented an experimental module for computing these correlations and adding them as an additional table in the L2 OIFITS files. The method is still under validation and is not described in this paper. An evolution of the model fitting tools is, of course, mandatory to fully exploit this.

\subsection{Propagation of uncertainties}
\label{sec:errpropa}
The error propagation from the raw to the calibrated visibility is divided into four parts:
\begin{itemize}
    \item The error of the raw visibility is statistically estimated by dividing an OB into ten sub-OBs and computing the RMS of the ten resulting visibilities.
    \item The transfer function is computed for each calibrator as the ratio of the raw visibility to the expected visibility knowing the angular diameter of the calibrator. So its error could be computed as the error of the ratio of two random variables. The method is presented in Appendix \ref{sec:ratio}.
    \item The transfer function at the time of science observation is the weighted average of the transfer functions of each calibrator taking into account the observing time of the calibrators as explained in \cite{lachaume2019}. The error of the interpolated transfer function is therefore given by the standard formula of the error of the weighted average.
    \item The calibrated visibility is computed as the ratio of the raw visibility by the interpolated transfer function. Therefore, its error is also computed with the method presented in the Appendix \ref{sec:ratio}. 
\end{itemize}

\section{Initial performance}
\label{sec:perfs}
The performance formalism of the SPICA-VIS concept has been presented in \cite{2017JOSAA..34A..37M} and revisited by \cite{friend}. It is unchanged, and we refer the reader to the equations detailed in this last paper. The first results of the commissioning allow us to review most of the numerical hypotheses used. Obviously, for the CHARA/SPICA instrument we consider the number of telescopes $N_\mathrm{tel}=6$. We focus our attention on the case of the low-resolution mode ($R=140$) for which the number of spectral channels is $N_\mathrm{SC}=60$. SPICA-VIS is using an ANDOR IXON 888 as the science detector. It should be noted that the amplification gain of the ANDOR detectors has to be recalibrated after few years of use. This operation maintains the best performance in the amplification of the signal before recording. The following values are considered: readout noise in electrons $Ron$=1, spurious Clock Induced Charges (CIC) $N_\mathrm{c}=0.0018$/pixel/frame, and dark events $N_\mathrm{d}=0.0002$/pixel/second.

\subsection{Photometry, transmission, and fiber coupling}
\label{sec:photometry}

\begin{figure}[ht]
   \center
\includegraphics[width=4.4cm]{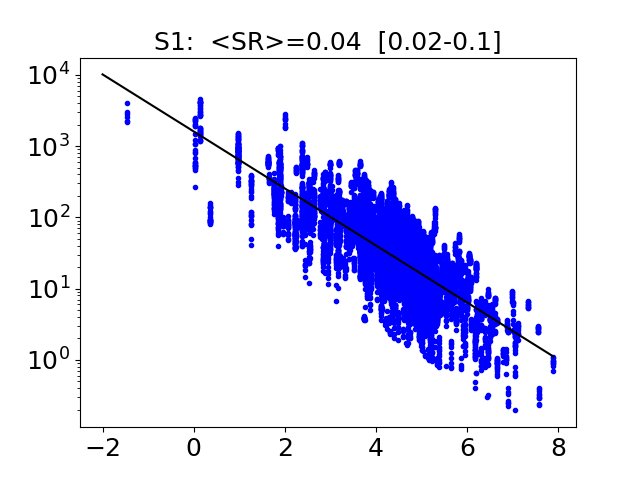}
\includegraphics[width=4.4cm]{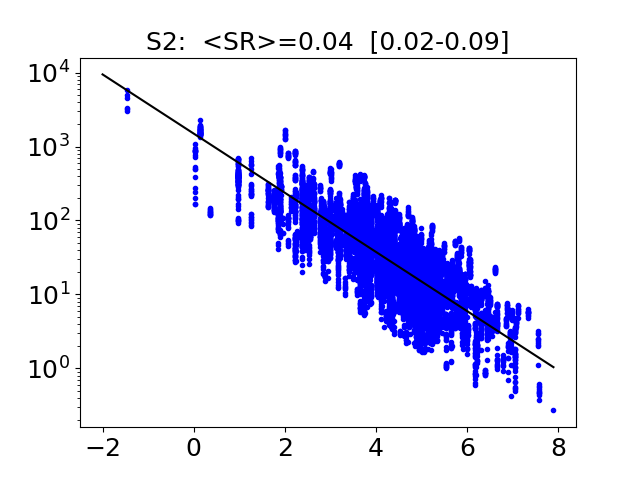}\\
\includegraphics[width=4.4cm]{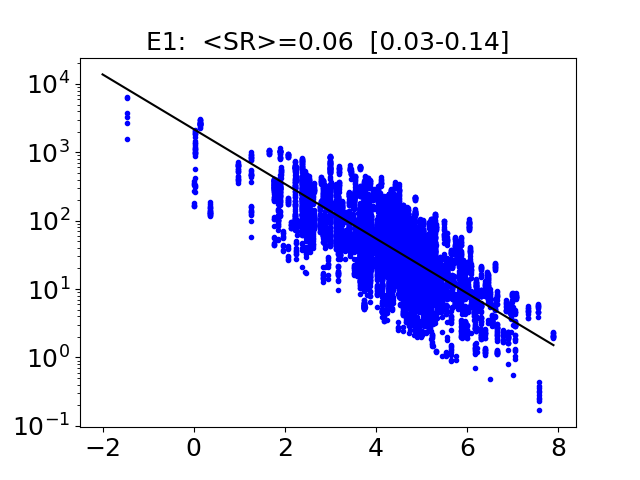}
\includegraphics[width=4.4cm]{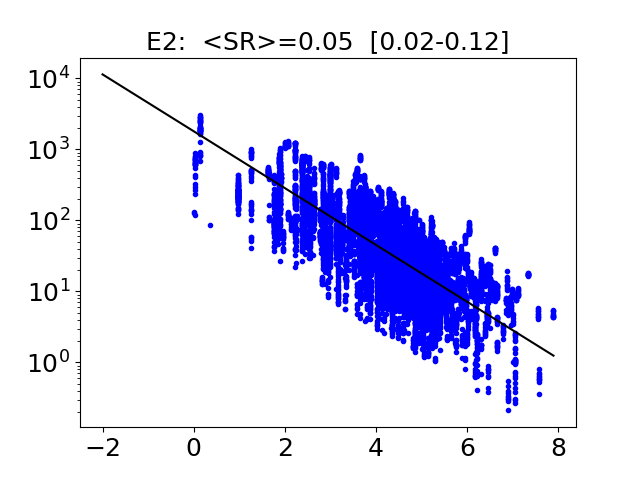}\\
\includegraphics[width=4.4cm]{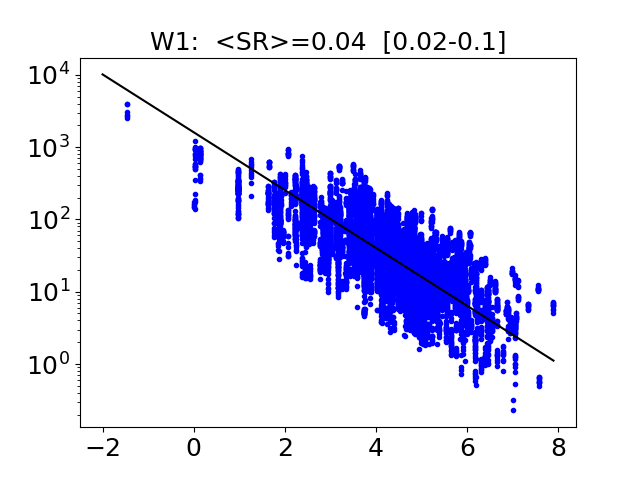}
\includegraphics[width=4.4cm]{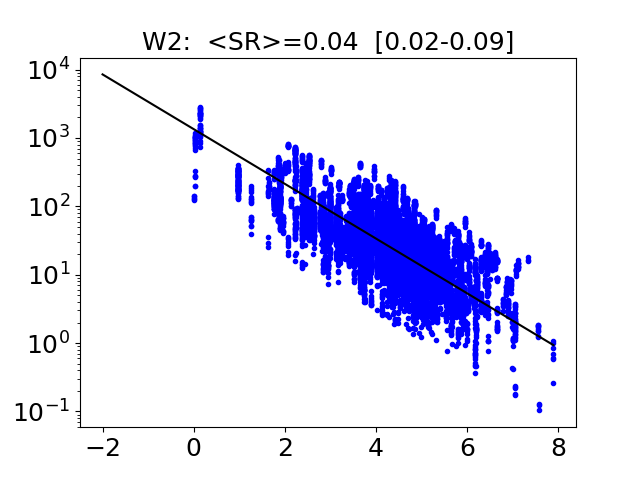}
   \caption{Statistics of number of photons per second recorded for each telescope (first row: S1-S2, second row E1-E2, and third row: W1-W2) in the photometric channel as a function of the magnitude of the star. A numerical model is overplotted following the standard law $10^{-0.4m}$, with $m$ the magnitude of the star. The scaling factor yields estimates of the mean Strehl ratio <SR> of each telescope, as well as the min and max values.}
   \label{fig:photometry}
\end{figure}

In Fig.~\ref{fig:photometry} we present the statistics of the flux recorded for each telescope as a function of the magnitude in the corresponding photometric channel of the instrument. For each telescope, the number of photons $N_\mathrm{p}$ for one frame could be written as:
\begin{multline}
	N_\mathrm{p} = T_\mathrm{C} T_\mathrm{AO} Co S_\mathrm{tel} \Delta\lambda DIT 10^{-0.4m} \Phi_\mathrm{0}Q
\end{multline}

Some parameters remain unchanged:
\begin{itemize}
	\item $T_\mathrm{AO} = 0.765$: transmission of adaptive optics,
	\item $S_\mathrm{tel}=0.75 m^2$: collecting area of the 1~m telescope,
	\item $\Delta \lambda=280 nm$: the spectral band,
	\item $DIT = 0.02 s$: detector integration time,
	\item $m$: the magnitude of the target,
	\item $\Phi_\mathrm{0} = 6.2 10^7 ph/m^2/nm/s$: the reference flux at $m=0$,
	\item $Q=0.9$: the quantum efficiency of the detector.
\end{itemize}
To correctly match the photometric measurements and their analysis presented in Fig.~\ref{fig:photometry}, we adjusted the overall transmission and the coupling efficiency :
\begin{itemize}
	\item $T_\mathrm{C} = 0.005$: instrumental transmission 
	\item $Co=0.02$: the coupling efficiency ($0.4 \times Strehl=0.05$)
\end{itemize}

Two important remarks must be made here. First, it can be seen that the instrumental transmission is lower than previously considered. This decrease in CHARA transmission could be due, for the data presented here, to the aging of the coating of the mirrors in the telescope. We expect an increase by at least a factor 2 with the fresh coating currently being made. Second, the coupling efficiency has been revisited because of the performance of adaptive optics that delivers a Strehl ratio \citep{Strehl} in the visible closer to 5\% (average value of the 6 telescopes) than the initially expected value of 25\%. However, it should be noted that some scopes (E1, E2, S1) can reach a Strehl Ratio above 10\%. Progress is being made on this topic, and observations demonstrate that all telescopes are now reaching a Strehl ratio between 12 and 15\% for median seeing conditions.

\subsection{Transfer function and estimation of the signal-to-noise ratio}
\label{sec:snr}

\begin{figure*}[htpb]
\sidecaption
  \centering
   \includegraphics[width=12cm,height=5cm]{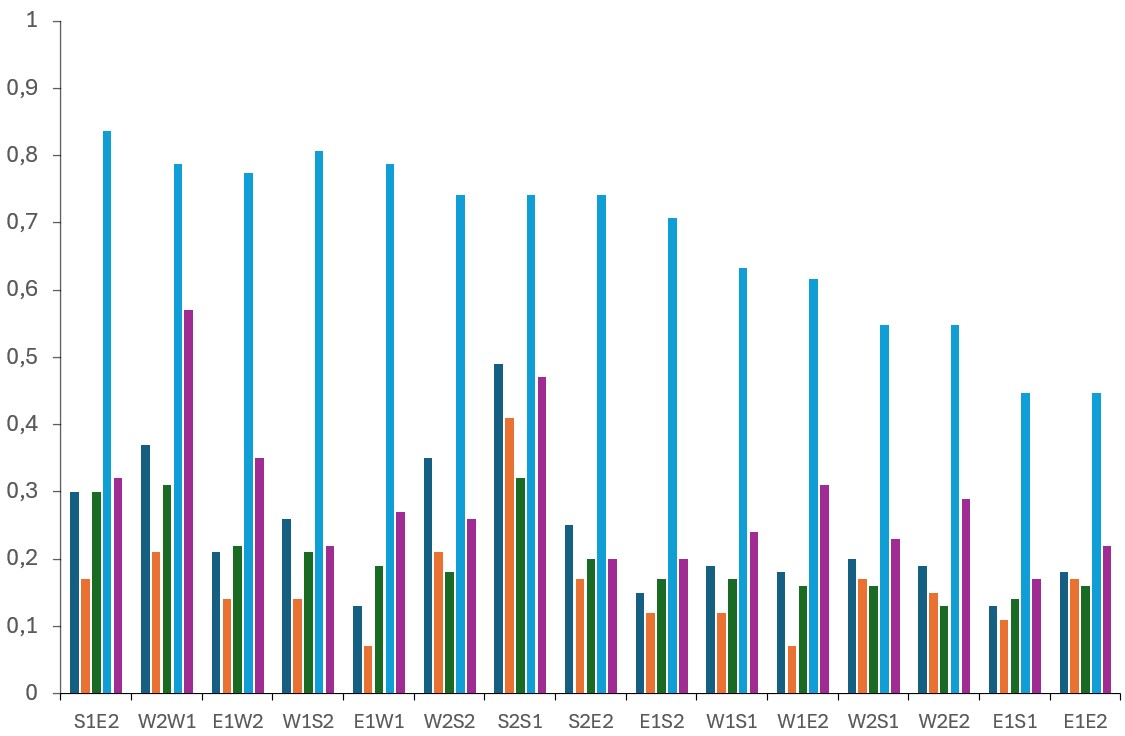}
   \caption{Comparison of the STS (light blue) and SKY transfer function as measured in 2023 (dark blue), 2024 (orange), 2025 (green), and 2026 (purple).}
   \label{fig:transfer}
\end{figure*}

In addition to the correct estimation of $N_p$, the estimation of the signal-to-noise ratio is highly dependent on the system visibility. On the STS, we have measured system visibilities $V_\mathrm{ins}$ between 0.85 and 0.50 depending almost linearly on the sampling of the fringes, encoded at different spatial frequencies on the detector (see Fig.\ref{fig:transfer}). However, the situation on sky has been very different during the first years of operations, with values ranging from 0.10 to 0.44, for an average visibility in the range $[680-850]$~nm. Fig.~\ref{fig:transfer} presents the comparison, baseline by baseline, between STS and the statistics of measurements in 2023, 2024 and 2025. \\
Defaults in the electronics of the control of the delay lines have strongly degraded the quality of the delay line tracking, generating spurious motions with an amplitude (up to 150-180~nm) depending of the speed of the cart. This issue was solved at the beginning of 2025, but it does not seem to be the main cause of reduction of the transfer function.\\ 
Baselines including E1 are the most affected by the degradation of the transfer function. E1 is also the cart that has the fastest motion to compensate for the rotation of the earth and often exhibits tracking errors at the level of $50-70$~nm. \\
But the main cause of degratation was identified at the beginning of 2026. The longitudinal dispersion (see Sect.~\ref{sec:ldc}) was in fact wrongly corrected, both in the infrared and in the visible. A numerical error was made in the air refraction index in the visible and has generated error in the calculation of the glass thickness. The consequence was first identified in the infrared, as important drifts of the group delay during the phase tracking by SPICA-FT, especially on the longest baselines, the most affected by the rapid changes in air path. For the visible, the consequence of this wrong correction was a large residual of dispersion. It was not identified because the estimation of the differential phase is done classically with the removal of a polynomial function to account for the residual optical path difference and dispersion. After the correct implementation of the dispersion, we have demonstrated improved performance of the fringe tracking and, even more importantly, better performance in the system visibility in the visible. Recent measurements demonstrate that we are now reaching, on sky, about 50\% of the system visibility in the STS source, as shown in Fig.~\ref{fig:transfer}. The 2026 statistics are not very rich for the moment and suffer from poor seeing in April and particularly low flux on South telescopes.\\
We also performed some SPICA recording at different DIT: $20, 13, 7, 5$~ms. On the S1S2 baselines, we obtain system visibility of $0.32, 0.36, 0.43, 0.49$ and $0.16, 0.19, 0.21, 0.26$ for the E1E2 baseline, which corresponds to a non negligible improvement of a factor 1.5. These experiments continue and involve fast reading modes with the ANDOR detectors.\\
As a conclusion, we can adopt a system visibility $V_\mathrm{sys}=0.35$ as an average value for the different baselines on sky. \\

In Fig.~\ref{fig:measuredsnr}, we present statistics of the measured signal-to-noise ratio of visibility as a function of the correlated flux for the three shortest baselines, S1S2, E1E2, and W1W2 for all observations for 2025. The noise model is superimposed on each plot with the corresponding system visibility (resp. 0.31, 0.10, and 0.31). Our noise model is capable of correctly reproducing the measurements with recently determined parameters. 

\begin{figure}[ht]
  \centering
   \includegraphics[width=8cm,height=5cm]{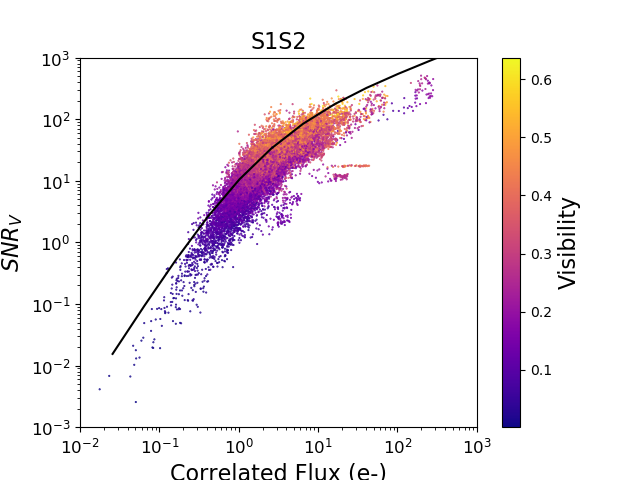}\\
   \includegraphics[width=8cm,height=5cm]{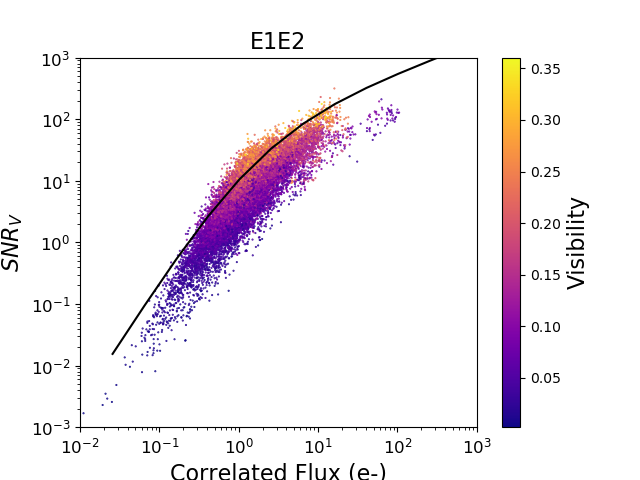}\\
   \includegraphics[width=8cm,height=5cm]{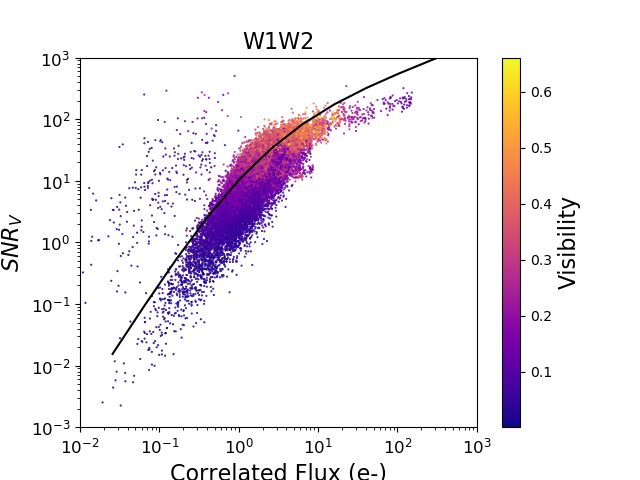}
   \caption{Distribution of the signal-to-noise ratio on visibility measured on SPICA as a function of the correlated flux and for the three shortest baselines, S1S2 at the top, E1E2 in the middle, and W1W2 at the bottom. The noise model is superimposed on each plot, taking into account the difference of system visibility (0.32/0.10/0.31 respectively for S1S2/E1E2/W1W2).}
   \label{fig:measuredsnr}
\end{figure}

Finally, in Figure~\ref{fig:snr2025} we present an estimate of the signal-to-noise ratio in low-resolution mode, for one individual spectral channel, a visibility V=1, and for 10~min of integration. The results are presented for three different DIT of 12, 200, and 2000~ms. In this simulation, we use the mean system visibility $V_\mathrm{sys}=0.5$, considering the new values after dispersion correction ($0.35$), the hypothesis of the shortest DIT, instrumental transmission $T_\mathrm C=0.01$, and a Strehl of 15\% that gives a coupling efficiency $Co=0.06$. With the recent improvements described in this section, we have been able to reach magnitude 6.5 under good seeing conditions, which seems to agree well with the numerical model presented. The efforts to improve coupling efficiency, overall transmission, and stabilization with SPICA-FT are critical to achieving our scientific goals.

\begin{figure}[ht]
  \centering
   \includegraphics[width=9.5cm]{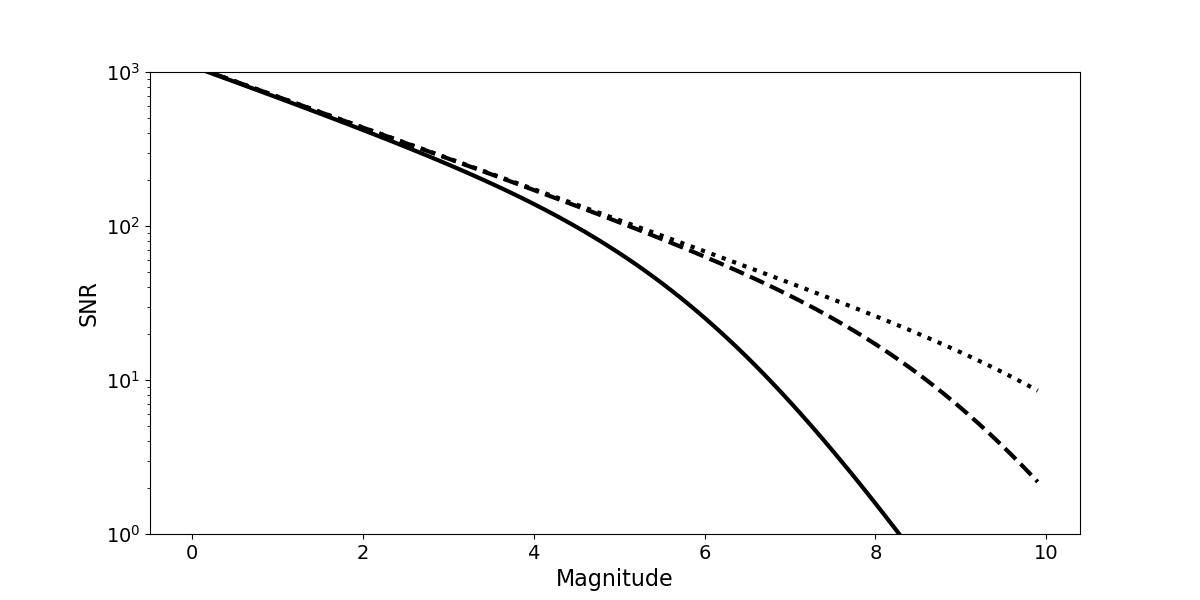}
   \caption{Signal-to-noise ratio per individual spectral channel for an integration of 10 min. Three different Detector Integration Time $[12, 200, 2000]~ms$ are considered and represented as [plain, dashed, dotted] lines.}
   \label{fig:snr2025}
\end{figure}

\section{Conclusion}
\label{sec:conclusion}
We have presented the recently commissioned visible 6T instrument for the CHARA Array, SPICA-VIS, and its complement SPICA-FT for the fringe tracking in the H band. Initially designed for a large and homogeneous survey of fundamental stellar parameters, SPICA is also open to a wide range of scientific programs. Observing time on SPICA is available to the community through the CHARA open-access program offered through NSF NOIRLab\footnote{https://chara.gsu.edu/observers/applying-for-chara-time}. Combined systematically with the infrared instruments MIRC-X and MYSTIC, this unique operating mode presents the opportunity for simultaneous observations over the very wide spectral coverage of the CHARA Array, from the R band to the J, H, and K bands. This multi-chromatic operation of CHARA is a very strong advantage for science programs.

After a detailed description of the instrument, we presented its operation, the data flow, and the details of the data-reduction pipeline. Despite a difficult start with multiple issues, we have finally succeeded in a better characterization of its performance and have established a solid numerical model of the system signal-to-noise ratio. 

First science papers are being published soon, addressing the main programs of the survey: a characterization of exoplanet host stars (Nowacki et al., in prep.), a calibration of seismic scaling relations (Vrard et al., in prep.), the surface brightness color relations for FGK stars \citep{ibanez2026}, a study of limb darkening across the HR diagram \citep{nayeem2026}, and stellar masses and binaries \citep{juraj2026}.  

\begin{acknowledgements}
This project has received funding from the European Research Council (ERC) under the European Union’s Horizon 2020 research and innovation programme (Grant agreement No. 101019653). SPICA has been funded by CNRS, Observatoire de la Côte d'Azur, Université Côte d'Azur, Région Sud, and the University of Aarhus. 
This work is based on observations obtained with the Georgia State University Center for High Angular Resolution Astronomy Array at Mount Wilson Observatory.  The CHARA Array is supported by the National Science Foundation under Grant No. AST-2034336 and AST-2407956. Institutional support has been provided from the GSU College of Arts and Sciences, Office of the Provost, and Office of the Vice President for Research and Economic Development.\\
This research has made use of the Jean-Marie Mariotti Center at https://jmmc.fr/.\\
S.K. acknowledges support from an ERC Consolidator Grant (Grant Agreement ID 101003096). MIRC-X has been build with funds from an ERC Starting Grant (Grant Agreement No.\ 639889) and an STFC equipment grant ST/X005143/1.\\
We are grateful to all colleagues and students who contributed to the SPICA observations and participated in regular progress meetings, with special thanks to Armando Domiciano da Souza, Orlagh Creevey, Manon Bailleul, Kamakshi Kaushik, Andrei Catalin-Popa, and Alexis Hellal for their invaluable support.\\
We warmly thank Becky Flores, Olli Majoinen, Heven Renteria, and Norm Vargas for their outstanding support during the night operations and Victor Castillo, Larry Webster, and Craig Woods for their invaluable assistance with the installation of SPICA and PSAUM, as well as during our various stays at Mount Wilson.
\end{acknowledgements}

\bibliographystyle{aa}
\bibliography{biblio}

@ARTICLE{Mourard2011,
       author = {{Mourard}, D. and {B{\'e}rio}, Ph. and {Perraut}, K. and {Ligi}, R. and {Blazit}, A. and {Clausse}, J.~M. and {Nardetto}, N. and {Spang}, A. and {Tallon-Bosc}, I. and {Bonneau}, D. and {Chesneau}, O. and {Delaa}, O. and {Millour}, F. and {Stee}, P. and {Le Bouquin}, J.~B. and {ten Brummelaar}, T. and {Farrington}, C. and {Goldfinger}, P.~J. and {Monnier}, J.~D.},
        title = "{Spatio-spectral encoding of fringes in optical long-baseline interferometry. Example of the 3T and 4T recombining mode of VEGA/CHARA}",
      journal = {\aap},
         year = 2011,
        month = jul,
       volume = {531},
          eid = {A110},
        pages = {A110},
          doi = {10.1051/0004-6361/201116976},
       adsurl = {https://ui.adsabs.harvard.edu/abs/2011A\&A...531A.110M}
}

@ARTICLE{Mourard2009,
       author = {{Mourard}, D. and {Clausse}, J.~M. and {Marcotto}, A. and {Perraut}, K. and {Tallon-Bosc}, I. and {B{\'e}rio}, Ph. and {Blazit}, A. and {Bonneau}, D. and {Bosio}, S. and {Bresson}, Y. and {Chesneau}, O. and {Delaa}, O. and {H{\'e}nault}, F. and {Hughes}, Y. and {Lagarde}, S. and {Merlin}, G. and {Roussel}, A. and {Spang}, A. and {Stee}, Ph. and {Tallon}, M. and {Antonelli}, P. and {Foy}, R. and {Kervella}, P. and {Petrov}, R. and {Thiebaut}, E. and {Vakili}, F. and {McAlister}, H. and {ten Brummelaar}, T. and {Sturmann}, J. and {Sturmann}, L. and {Turner}, N. and {Farrington}, C. and {Goldfinger}, P.~J.},
        title = "{VEGA: Visible spEctroGraph and polArimeter for the CHARA array: principle and performance}",
      journal = {\aap},
         year = 2009,
        month = dec,
       volume = {508},
       number = {2},
        pages = {1073-1083},
          doi = {10.1051/0004-6361/200913016},
       adsurl = {https://ui.adsabs.harvard.edu/abs/2009A\&A...508.1073M}
}

@ARTICLE{gi2t,
       author = {{Mourard}, D. and {Tallon-Bosc}, I. and {Blazit}, A. and {Bonneau}, D. and {Merlin}, G. and {Morand}, F. and {Vakili}, F. and {Labeyrie}, A.},
        title = "{The GI2T interferometer on Plateau de Calern}",
      journal = {\aap},
         year = 1994,
        month = mar,
       volume = {283},
       number = {2},
        pages = {705-713},
       adsurl = {https://ui.adsabs.harvard.edu/abs/1994A\&A...283..705M}
}

@ARTICLE{mircx,
       author = {{Anugu}, Narsireddy and {Le Bouquin}, Jean-Baptiste and {Monnier}, John D. and {Kraus}, Stefan and {Setterholm}, Benjamin R. and {Labdon}, Aaron and {Davies}, Claire L. and {Lanthermann}, Cyprien and {Gardner}, Tyler and {Ennis}, Jacob and {Johnson}, Keith J.~C. and {Ten Brummelaar}, Theo and {Schaefer}, Gail and {Sturmann}, Judit},
        title = "{MIRC-X: A Highly Sensitive Six-telescope Interferometric Imager at the CHARA Array}",
      journal = {The Astronomical Journal},
         year = 2020,
        month = oct,
       volume = {160},
       number = {4},
          eid = {158},
        pages = {158},
          doi = {10.3847/1538-3881/aba957},
archivePrefix = {arXiv},
       eprint = {2007.12320},
 primaryClass = {astro-ph.IM},
       adsurl = {https://ui.adsabs.harvard.edu/abs/2020AJ....160..158A}
}

@ARTICLE{ldc,
       author = {{Pannetier}, Cyril and {Mourard}, Denis and {Cassaing}, Fr{\'e}d{\'e}ric and {Lagarde}, St{\'e}phane and {Le Bouquin}, Jean-Baptiste and {Monnier}, John and {Sturmann}, Judit and {Ten Brummelaar}, Theo},
        title = "{Compensation of differential dispersion: application to multiband stellar interferometry}",
      journal = {MNRAS},
         year = 2021,
        month = oct,
       volume = {507},
       number = {1},
        pages = {1369-1380},
          doi = {10.1093/mnras/stab1922},
archivePrefix = {arXiv},
       eprint = {2109.07163},
 primaryClass = {astro-ph.IM},
       adsurl = {https://ui.adsabs.harvard.edu/abs/2021MNRAS.507.1369P}
}

@ARTICLE{lacour,
       author = {{Lacour}, S. and {Dembet}, R. and {Abuter}, R. and {F{\'e}dou}, P. and {Perrin}, G. and {Choquet}, {\'E}. and {Pfuhl}, O. and {Eisenhauer}, F. and {Woillez}, J. and {Cassaing}, F. and {Wieprecht}, E. and {Ott}, T. and {Wiezorrek}, E. and {Tristram}, K.~R.~W. and {Wolff}, B. and {Ram{\'\i}rez}, A. and {Haubois}, X. and {Perraut}, K. and {Straubmeier}, C. and {Brandner}, W. and {Amorim}, A.},
        title = "{The GRAVITY fringe tracker}",
      journal = {A\&A},
         year = 2019,
        month = apr,
       volume = {624},
          eid = {A99},
        pages = {A99},
          doi = {10.1051/0004-6361/201834981},
archivePrefix = {arXiv},
       eprint = {1901.03202},
 primaryClass = {astro-ph.IM},
       adsurl = {https://ui.adsabs.harvard.edu/abs/2019A\&A...624A..99L}
}

@ARTICLE{mystic,
       author = {{Setterholm}, Benjamin R. and {Monnier}, John D. and {Le Bouquin}, Jean-Baptiste and {Anugu}, Narsireddy and {Ennis}, Jacob and {Jocou}, Laurent and {Ibrahim}, Nour and {Kraus}, Stefan and {Anderson}, Matthew D. and {Chhabra}, Sorabh and {Codron}, Isabelle and {Farrington}, Christopher D. and {Flores}, Becky and {Gardner}, Tyler and {Gutierrez}, Mayra and {Lanthermann}, Cyprien and {Majoinen}, Olli W. and {Mortimer}, Daniel J. and {Schaefer}, Gail and {Scott}, Nicholas J. and {ten Brummelaar}, Theo and {Vargas}, Norman L.},
        title = "{MYSTIC: a high angular resolution K-band imager at CHARA}",
      journal = {Journal of Astronomical Telescopes, Instruments, and Systems},
         year = 2023,
        month = apr,
       volume = {9},
          eid = {025006},
        pages = {025006},
          doi = {10.1117/1.JATIS.9.2.025006},
       adsurl = {https://ui.adsabs.harvard.edu/abs/2023JATIS...9b5006S}
}

@ARTICLE{labeyrie74,
   author = {{Labeyrie}, A.},
    title = "{Interference fringes obtained on VEGA with two optical telescopes}",
  journal = {ApJ},
     year = 1975,
    month = mar,
   volume = 196,
    pages = {L71-L75},
      doi = {10.1086/181747},
   adsurl = {http://adsabs.harvard.edu/abs/1975ApJ...196L..71L}
}

@ARTICLE{mark3,
   author = {{Shao}, M. and {Colavita}, M.~M. and {Hines}, B.~E. and {Staelin}, D.~H. and
	{Hutter}, D.~J. and {Johnston}, K.~J. and {Mozurkewich}, D. and
	{Simon}, R.~S. and {Hershey}, J.~L. and {Hughes}, J.~A. and
	{Kaplan}, G.~H.},
    title = "{The Mark III stellar interferometer}",
  journal = {A\&A},
     year = 1988,
    month = mar,
   volume = 193,
    pages = {357-371},
   adsurl = {http://adsabs.harvard.edu/abs/1988A%26A...193..357S}
}

@ARTICLE{susi,
   author = {{Davis}, J. and {Tango}, W.~J.},
    title = "{The Sydney University 11.4 M prototype stellar interferometer}",
  journal = {Proceedings of the Astronomical Society of Australia},
     year = 1985,
   volume = 6,
    pages = {34-38},
   adsurl = {http://adsabs.harvard.edu/abs/1985PASAu...6...34D}
}

@ARTICLE{coast,
   author = {{Baldwin}, J.~E. and {Beckett}, M.~G. and {Boysen}, R.~C. and
	{Burns}, D. and {Buscher}, D.~F. and {Cox}, G.~C. and {Haniff}, C.~A. and
	{Mackay}, C.~D. and {Nightingale}, N.~S. and {Rogers}, J. and
	{Scheuer}, P.~A.~G. and {Scott}, T.~R. and {Tuthill}, P.~G. and
	{Warner}, P.~J. and {Wilson}, D.~M.~A. and {Wilson}, R.~W.},
    title = "{The first images from an optical aperture synthesis array: mapping of Capella with COAST at two epochs.}",
  journal = {A\&A},
     year = 1996,
    month = feb,
   volume = 306,
    pages = {L13},
   adsurl = {http://adsabs.harvard.edu/abs/1996A%26A...306L..13B}
}

@ARTICLE{npoi,
   author = {{Armstrong}, J.~T. and {Mozurkewich}, D. and {Rickard}, L.~J. and
	{Hutter}, D.~J. and {Benson}, J.~A. and {Bowers}, P.~F. and
	{Elias}, II, N.~M. and {Hummel}, C.~A. and {Johnston}, K.~J. and
	{Buscher}, D.~F. and {Clark}, III, J.~H. and {Ha}, L. and {Ling}, L.-C. and
	{White}, N.~M. and {Simon}, R.~S.},
    title = "{The Navy Prototype Optical Interferometer}",
  journal = {ApJ},
     year = 1998,
    month = mar,
   volume = 496,
    pages = {550-571},
      doi = {10.1086/305365},
   adsurl = {http://adsabs.harvard.edu/abs/1998ApJ...496..550A}
}

@ARTICLE{chara,
   author = {{ten Brummelaar}, T.~A. and {McAlister}, H.~A. and {Ridgway}, S.~T. and
	{Bagnuolo}, Jr., W.~G. and {Turner}, N.~H. and {Sturmann}, L. and
	{Sturmann}, J. and {Berger}, D.~H. and {Ogden}, C.~E. and {Cadman}, R. and
	{Hartkopf}, W.~I. and {Hopper}, C.~H. and {Shure}, M.~A.},
    title = "{First Results from the CHARA Array. II. A Description of the Instrument}",
  journal = {ApJ},
   eprint = {arXiv:astro-ph/0504082},
     year = 2005,
    month = jul,
   volume = 628,
    pages = {453-465},
      doi = {10.1086/430729},
   adsurl = {http://cdsads.u-strasbg.fr/abs/2005ApJ...628..453T}
}

@INPROCEEDINGS{charaao1,
   author = {{ten Brummelaar}, T. and {Che}, X. and {McAlister}, H. and {Ireland}, M. and
	{Monnier}, J. and {Mourard}, D. and {Ridgway}, S. and {Sturmann}, J. and
	{Sturmann}, L. and {Turner}, N. and {Tuthill}, P.},
    title = "{CHARA array adaptive optics II: non-common-path correction and downstream optics}",
booktitle = {Adaptive Optics Systems IV},
     year = 2014,
   series = {Presented at the Society of Photo-Optical Instrumentation Engineers (SPIE) Conference},
   volume = 9148,
    month = aug,
      eid = {91484Q},
    pages = {91484Q},
      doi = {10.1117/12.2056662},
   adsurl = {http://adsabs.harvard.edu/abs/2014SPIE.9148E..4QT}
}

@INPROCEEDINGS{charaao2,
   author = {{Che}, X. and {Sturmann}, L. and {Monnier}, J.~D. and {ten Brummelaar}, T.~A. and
	{Sturmann}, J. and {Ridgway}, S.~T. and {Ireland}, M.~J. and
	{Turner}, N.~H. and {McAlister}, H.~A.},
    title = "{The CHARA array adaptive optics I: common-path optical and mechanical design, and preliminary on-sky results}",
booktitle = {Adaptive Optics Systems IV},
     year = 2014,
   series = {Presented at the Society of Photo-Optical Instrumentation Engineers (SPIE) Conference},
   volume = 9148,
    month = jul,
      eid = {914830},
    pages = {914830},
      doi = {10.1117/12.2055693},
   adsurl = {http://adsabs.harvard.edu/abs/2014SPIE.9148E..30C}
}

@ARTICLE{vision,
   author = {{Garcia}, E.~V. and {Muterspaugh}, M.~W. and {van Belle}, G. and
	{Monnier}, J.~D. and {Stassun}, K.~G. and {Ghasempour}, A. and
	{Clark}, J.~H. and {Zavala}, R.~T. and {Benson}, J.~A. and {Hutter}, D.~J. and
	{Schmitt}, H.~R. and {Baines}, E.~K. and {Jorgensen}, A.~M. and
	{Strosahl}, S.~G. and {Sanborn}, J. and {Zawicki}, S.~J. and
	{Sakosky}, M.~F. and {Swihart}, S.},
    title = "{Vision: A Six-telescope Fiber-fed Visible Light Beam Combiner for the Navy Precision Optical Interferometer}",
  journal = {\pasp},
archivePrefix = "arXiv",
   eprint = {1601.00036},
 primaryClass = "astro-ph.IM",
     year = 2016,
    month = may,
   volume = 128,
   number = 5,
    pages = {055004},
      doi = {10.1088/1538-3873/128/963/055004},
   adsurl = {http://adsabs.harvard.edu/abs/2016PASP..128e5004G}
}

@ARTICLE{stee2017,
   author = {{Stee}, P. and {Allard}, F. and {Benisty}, M. and {Bigot}, L. and 
	{Blind}, N. and {Boffin}, H. and {Borges Fernandes}, M. and 
	{Carciofi}, A. and {Chiavassa}, A. and {Creevey}, O. and {Cruzalebes}, P. and 
	{de Wit}, W.-J. and {Domiciano de Souza}, A. and {Elvis}, M. and 
	{Fabas}, N. and {Faes}, D. and {Gallenne}, A. and {Guerrero Pena}, C. and 
	{Hillen}, M. and {Hoenig}, S. and {Ireland}, M. and {Kervella}, P. and 
	{Kishimoto}, M. and {Kostogryz}, N. and {Kraus}, S. and {Labeyrie}, A. and 
	{Le Bouquin}, J.-B. and {Lebre}, A. and {Ligi}, R. and {Marconi}, A. and 
	{Marsh}, T. and {Meilland}, A. and {Millour}, F. and {Monnier}, J. and 
	{Mourard}, D. and {Nardetto}, N. and {Ohnaka}, K. and {Paladini}, C. and 
	{Perraut}, K. and {Perrin}, G. and {Petit}, P. and {Petrov}, R. and 
	{Rakshit}, S. and {Schaefer}, G. and {Schneider}, J. and {Shulyak}, D. and 
	{Simon}, M. and {Soulez}, F. and {Steeghs}, D. and {Tallon-Bosc}, I. and 
	{Tallon}, M. and {ten Brummelaar}, T. and {Thiebaut}, E. and 
	{Th{\'e}venin}, F. and {Van Winckel}, H. and {Wittkowski}, M. and 
	{Zorec}, J.},
    title = "{Science cases for a visible interferometer}",
  journal = {ArXiv e-prints},
archivePrefix = "arXiv",
   eprint = {1703.02395},
 primaryClass = "astro-ph.SR",
     year = 2017,
    month = mar,
   adsurl = {http://adsabs.harvard.edu/abs/2017arXiv170302395S}
}

@ARTICLE{vlti,
       author = {{Woillez}, J. and {Gont{\'e}}, F. and {Abad}, J.~A. and {Abadie}, S. and {Abuter}, R. and {Accardo}, M. and {Acu{\~n}a}, M. and {Alonso}, J. and {Andolfato}, L. and {Avila}, G. and {Barriga}, P.~J. and {Beltran}, J. and {Berger}, J. -P. and {Bollados}, C. and {Bourget}, P. and {Brast}, R. and {Bristow}, P. and {Caniguante}, L. and {Castillo}, R. and {Conzelmann}, R. and {Cortes}, A. and {Delplancke}, F. and {Dell Valle}, D. and {Derie}, F. and {Diaz}, A. and {Donoso}, R. and {Duhoux}, Ph. and {Dupuy}, C. and {Elao}, C. and {Egner}, S. and {Fuenteseca}, E. and {Fernandez}, R. and {Gaytan}, D. and {Glindemann}, A. and {Gonzales}, J. and {Guisard}, S. and {Hagenauer}, P. and {Haimerl}, A. and {Heinz}, V. and {Henriquez}, J.~P. and {van der Heyden}, P. and {Hubin}, N. and {Huerta}, R. and {Jochum}, L. and {Kirchbauer}, J. -P. and {Leiva}, A. and {L{\'e}v{\^e}que}, S. and {Lizon}, J. -P. and {Luco}, F. and {Mardones}, P. and {Mellado}, A. and {M{\'e}rand}, A. and {Osorio}, J. and {Ott}, J. and {Pallanca}, L. and {Pavez}, M. and {Pasquini}, L. and {Percheron}, I. and {Pirard}, J. -F. and {Phan}, D.~T. and {Pineda}, J.~C. and {Pino}, A. and {Poupar}, S. and {Ram{\'\i}rez}, A. and {Reinero}, C. and {Riquelme}, M. and {Romero}, J. and {Rivinius}, Th. and {Rojas}, C. and {Rozas}, F. and {Salgado}, F. and {Sch{\"o}ller}, M. and {Schuhler}, N. and {Siclari}, W. and {Stephan}, C. and {Tamblay}, R. and {Tapia}, M. and {Tristram}, K. and {Valdes}, G. and {de Wit}, W. -J. and {Wright}, A. and {Zins}, G.},
        title = "{VLTI: First Light for the Second Generation}",
      journal = {The Messenger},
         year = 2015,
        month = dec,
       volume = {162},
        pages = {16-18},
       adsurl = {https://ui.adsabs.harvard.edu/abs/2015Msngr.162...16W}
}

@ARTICLE{abaur,
       author = {{Rousselet-Perraut}, K. and {Benisty}, M. and {Mourard}, D. and {Rajabi}, S. and {Bacciotti}, F. and {B{\'e}rio}, Ph. and {Bonneau}, D. and {Chesneau}, O. and {Clausse}, J.~M. and {Delaa}, O. and {Marcotto}, A. and {Roussel}, A. and {Spang}, A. and {Stee}, Ph. and {Tallon-Bosc}, I. and {McAlister}, H. and {ten Brummelaar}, T. and {Sturmann}, J. and {Sturmann}, L. and {Turner}, N. and {Farrington}, C. and {Goldfinger}, P.~J.},
        title = "{The H{\ensuremath{\alpha}} line forming region of AB Aurigae spatially resolved at sub-AU with the VEGA/CHARA spectro-interferometer}",
      journal = {\aap},
         year = 2010,
        month = jun,
       volume = {516},
          eid = {L1},
        pages = {L1},
          doi = {10.1051/0004-6361/201014720},
archivePrefix = {arXiv},
       eprint = {1005.5267},
 primaryClass = {astro-ph.SR},
       adsurl = {https://ui.adsabs.harvard.edu/abs/2010A\&A...516L...1R}
}

@ARTICLE{bigot,
       author = {{Bigot}, L. and {Mourard}, D. and {Berio}, P. and {Th{\'e}venin}, F. and {Ligi}, R. and {Tallon-Bosc}, I. and {Chesneau}, O. and {Delaa}, O. and {Nardetto}, N. and {Perraut}, K. and {Stee}, Ph. and {Boyajian}, T. and {Morel}, P. and {Pichon}, B. and {Kervella}, P. and {Schmider}, F.~X. and {McAlister}, H. and {ten Brummelaar}, T. and {Ridgway}, S.~T. and {Sturmann}, J. and {Sturmann}, L. and {Turner}, N. and {Farrington}, C. and {Goldfinger}, P.~J.},
        title = "{The diameter of the CoRoT target HD 49933. Combining the 3D limb darkening, asteroseismology, and interferometry}",
      journal = {\aap},
         year = 2011,
        month = oct,
       volume = {534},
          eid = {L3},
        pages = {L3},
          doi = {10.1051/0004-6361/201117349},
archivePrefix = {arXiv},
       eprint = {1110.0985},
 primaryClass = {astro-ph.SR},
       adsurl = {https://ui.adsabs.harvard.edu/abs/2011A\&A...534L...3B}
}

@ARTICLE{ligi,
       author = {{Ligi}, R. and {Creevey}, O. and {Mourard}, D. and {Crida}, A. and {Lagrange}, A. -M. and {Nardetto}, N. and {Perraut}, K. and {Schultheis}, M. and {Tallon-Bosc}, I. and {ten Brummelaar}, T.},
        title = "{Radii, masses, and ages of 18 bright stars using interferometry and new estimations of exoplanetary parameters}",
      journal = {\aap},
         year = 2016,
        month = feb,
       volume = {586},
          eid = {A94},
        pages = {A94},
          doi = {10.1051/0004-6361/201527054},
archivePrefix = {arXiv},
       eprint = {1511.03197},
 primaryClass = {astro-ph.SR},
       adsurl = {https://ui.adsabs.harvard.edu/abs/2016A\&A...586A..94L}
}

@ARTICLE{phiper,
       author = {{Mourard}, D. and {Monnier}, J.~D. and {Meilland}, A. and {Gies}, D. and {Millour}, F. and {Benisty}, M. and {Che}, X. and {Grundstrom}, E.~D. and {Ligi}, R. and {Schaefer}, G. and {Baron}, F. and {Kraus}, S. and {Zhao}, M. and {Pedretti}, E. and {Berio}, P. and {Clausse}, J.~M. and {Nardetto}, N. and {Perraut}, K. and {Spang}, A. and {Stee}, P. and {Tallon-Bosc}, I. and {McAlister}, H. and {ten Brummelaar}, T. and {Ridgway}, S.~T. and {Sturmann}, J. and {Sturmann}, L. and {Turner}, N. and {Farrington}, C.},
        title = "{Spectral and spatial imaging of the Be+sdO binary <ASTROBJ>{\ensuremath{\phi}} Persei</ASTROBJ>}",
      journal = {\aap},
         year = 2015,
        month = may,
       volume = {577},
          eid = {A51},
        pages = {A51},
          doi = {10.1051/0004-6361/201425141},
archivePrefix = {arXiv},
       eprint = {1503.03423},
 primaryClass = {astro-ph.SR},
       adsurl = {https://ui.adsabs.harvard.edu/abs/2015A\&A...577A..51M}
}

@ARTICLE{sbcr,
       author = {{Salsi}, A. and {Nardetto}, N. and {Mourard}, D. and {Graczyk}, D. and {Taormina}, M. and {Creevey}, O. and {Hocd{\'e}}, V. and {Morand}, F. and {Perraut}, K. and {Pietrzynski}, G. and {Schaefer}, G.~H.},
        title = "{Progress on the calibration of surface brightness-color relations for early- and late-type stars}",
      journal = {\aap},
         year = 2021,
        month = aug,
       volume = {652},
          eid = {A26},
        pages = {A26},
          doi = {10.1051/0004-6361/202140763},
archivePrefix = {arXiv},
       eprint = {2106.01073},
 primaryClass = {astro-ph.SR},
       adsurl = {https://ui.adsabs.harvard.edu/abs/2021A\&A...652A..26S}
}

@INPROCEEDINGS{spie2012,
       author = {{Mourard}, Denis and {Challouf}, Mounir and {Ligi}, Roxanne and {B{\'e}rio}, Philippe and {Clausse}, Jean-Michel and {Gerakis}, J{\'e}r{\^o}me and {Bourges}, Laurent and {Nardetto}, Nicolas and {Perraut}, Karine and {Tallon-Bosc}, Isabelle and {McAlister}, H. and {ten Brummelaar}, T. and {Ridgway}, S. and {Sturmann}, J. and {Sturmann}, L. and {Turner}, N. and {Farrington}, C. and {Goldfinger}, P.~J.},
        title = "{Performance, results, and prospects of the visible spectrograph VEGA on CHARA}",
    booktitle = {Optical and Infrared Interferometry III},
         year = 2012,
       editor = {{Delplancke}, Fran{\c{c}}oise and {Rajagopal}, Jayadev K. and {Malbet}, Fabien},
       series = {Society of Photo-Optical Instrumentation Engineers (SPIE) Conference Series},
       volume = {8445},
        month = jul,
          eid = {84450K},
        pages = {84450K},
          doi = {10.1117/12.925223},
       adsurl = {https://ui.adsabs.harvard.edu/abs/2012SPIE.8445E..0KM}
}

@ARTICLE{friend,
       author = {{Martinod}, M.~A. and {Mourard}, D. and {B{\'e}rio}, P. and {Perraut}, K. and {Meilland}, A. and {Bailet}, C. and {Bresson}, Y. and {ten Brummelaar}, T. and {Clausse}, J.~M. and {Dejonghe}, J. and {Ireland}, M. and {Millour}, F. and {Monnier}, J.~D. and {Sturmann}, J. and {Sturmann}, L. and {Tallon}, M.},
        title = "{Fibered visible interferometry and adaptive optics: FRIEND at CHARA}",
      journal = {\aap},
         year = 2018,
        month = oct,
       volume = {618},
          eid = {A153},
        pages = {A153},
          doi = {10.1051/0004-6361/201731386},
       adsurl = {https://ui.adsabs.harvard.edu/abs/2018A\&A...618A.153M}
}

@article{smfib,
	author = {{Perrin}, G. and {Jocou, L.} and {Perraut, K.} and {Berger, J.-Ph.} and {Dembet, R.} and {Fédou, P.} and {Lacour, S.} and {Chapron, F.} and {Collin, C.} and {Poulain, S.} and {Cardin, V.} and {Joulain, F.} and {Eisenhauer, F.} and {Haubois, X.} and {Gillessen, S.} and {Haug, M.} and {Hausmann, F.} and {Kervella, P.} and {Léna, P.} and {Lippa, M.} and {Pfuhl, O.} and {Rabien, S.} and {Amorim, A.} and {Brandner, W.} and {Straubmeier, C.}},
	title = {Single-mode waveguides for GRAVITY - II. Single-mode fibers and Fiber Control Unit},
	DOI= "10.1051/0004-6361/202347587",
	url= "https://doi.org/10.1051/0004-6361/202347587",
	journal = {A\&A},
	year = 2024,
	volume = 681,
	pages = "A26",
}

@ARTICLE{2017JOSAA..34A..37M,
       author = {{Mourard}, Denis and {B{\'e}rio}, Philippe and {Perraut}, Karine and {Clausse}, Jean-Michel and {Creevey}, Orlagh and {Martinod}, Marc-Antoine and {Meilland}, Anthony and {Millour}, Florentin and {Nardetto}, Nicolas},
        title = "{SPICA, Stellar Parameters and Images with a Cophased Array: a 6T visible combiner for the CHARA array}",
      journal = {Journal of the Optical Society of America A},
         year = 2017,
        month = may,
       volume = {34},
       number = {5},
        pages = {A37},
          doi = {10.1364/JOSAA.34.000A37},
       adsurl = {https://ui.adsabs.harvard.edu/abs/2017JOSAA..34A..37M}
}

@INPROCEEDINGS{spica2024,
       author = {{Mourard}, Denis and {Meilland}, Anthony and {Iba{\~n}ez Bustos}, Romina and {Jonak}, Juraj and {Berio}, Philippe and {Dejonghe}, Julien and {Lecron}, Daniel and {Morand}, Fr{\'e}d{\'e}ric and {Salabert}, David and {Allouche}, Fatm{\'e} and {Anugu}, Narsireddy and {Bosio}, Sandra and {Bourges}, Laurent and {Creevey}, Orlagh and {Deheuvels}, S{\'e}bastien and {Domiciano de Souza}, Armando and {Ebrahimkutty}, Nayeem and {Gies}, Doug R. and {Kubiak}, Karolina and {Ligi}, Roxanne and {Ligon}, Robert and {Mella}, Guillaume and {Nardetto}, Nicolas and {Perraut}, Karine and {Pitiot}, Christophe and {Rousseau}, Sylvain and {Vrard}, Mathieu and {Schaefer}, Gail H. and {Spang}, Alain and {Turner}, Nils and {Wittkowski}, Markus and {Zumbo}, Florian},
        title = "{CHARA/SPICA: the new 6T visible combiner for the CHARA Array}",
    booktitle = {Optical and Infrared Interferometry and Imaging IX},
         year = 2024,
       editor = {{Kammerer}, Jens and {Sallum}, Stephanie and {Sanchez-Bermudez}, Joel},
       series = {Society of Photo-Optical Instrumentation Engineers (SPIE) Conference Series},
       volume = {13095},
        month = aug,
          eid = {1309503},
        pages = {1309503},
          doi = {10.1117/12.3019970},
       adsurl = {https://ui.adsabs.harvard.edu/abs/2024SPIE13095E..03M}
}

@INPROCEEDINGS{spica2022,
       author = {{Mourard}, Denis and {Berio}, Philippe and {Pannetier}, Cyril and {Nardetto}, Nicolas and {Allouche}, Fatm{\'e} and {Bailet}, Christophe and {Dejonghe}, Julien and {Geneslay}, Pierre and {Jacqmart}, Estelle and {Lagarde}, St{\'e}phane and {Lecron}, Daniel and {Morand}, Fr{\'e}d{\'e}ric and {Rousseau}, Sylvain and {Salabert}, David and {Spang}, Alain and {Albrecht}, Simon and {Anugu}, Narsireddy and {Bourg{\`e}s}, Laurent and {ten Brummelaar}, Theo A. and {Creevey}, Orlagh and {Deheuvels}, Sebastien and {Domiciano de Souza}, Armando and {Gies}, Doug and {Ligi}, Roxanne and {Mella}, Guillaume and {Perraut}, Karine and {Schaefer}, Gail and {Wittkowski}, Markus},
        title = "{CHARA/SPICA: a six-telescope visible instrument for the CHARA Array}",
    booktitle = {Optical and Infrared Interferometry and Imaging VIII},
         year = 2022,
       editor = {{M{\'e}rand}, Antoine and {Sallum}, Stephanie and {Sanchez-Bermudez}, Joel},
       series = {Society of Photo-Optical Instrumentation Engineers (SPIE) Conference Series},
       volume = {12183},
        month = aug,
          eid = {1218308},
        pages = {1218308},
          doi = {10.1117/12.2628881},
archivePrefix = {arXiv},
       eprint = {2210.09096},
 primaryClass = {astro-ph.IM},
       adsurl = {https://ui.adsabs.harvard.edu/abs/2022SPIE12183E..08M}
}

@ARTICLE{PLATO,
       author = {{Rauer}, Heike and {Aerts}, Conny and {Cabrera}, Juan and {Deleuil}, Magali and {Erikson}, Anders and {Gizon}, Laurent and {Goupil}, Mariejo and {Heras}, Ana and {Walloschek}, Thomas and {Lorenzo-Alvarez}, Jose and {Marliani}, Filippo and {Martin-Garcia}, C{\'e}sar and {Mas-Hesse}, J. Miguel and {O'Rourke}, Laurence and {Osborn}, Hugh and {Pagano}, Isabella and {Piotto}, Giampaolo and {Pollacco}, Don and {Ragazzoni}, Roberto and {Ramsay}, Gavin and {Udry}, St{\'e}phane and {Appourchaux}, Thierry and {Benz}, Willy and {Brandeker}, Alexis and {G{\"u}del}, Manuel and {Janot-Pacheco}, Eduardo and {Kabath}, Petr and {Kjeldsen}, Hans and {Min}, Michiel and {Santos}, Nuno and {Smith}, Alan and {Suarez}, Juan-Carlos and {Werner}, Stephanie C. and {Aboudan}, Alessio and {Abreu}, Manuel and {Acu{\~n}a}, Lorena and {Adams}, Moritz and {Adibekyan}, Vardan and {Affer}, Laura and {Agneray}, Fran{\c{c}}ois and {Agnor}, Craig and {Aguirre B{\o}rsen-Koch}, Victor and {Ahmed}, Saad and {Aigrain}, Suzanne and {Al-Bahlawan}, Ashraf and {Alcacera Gil}, Ma de los Angeles and {Alei}, Eleonora and {Alencar}, Silvia and {Alexander}, Richard and {Alfonso-Garz{\'o}n}, Julia and {Alibert}, Yann and {Allende Prieto}, Carlos and {Almeida}, Leonardo and {Alonso Sobrino}, Roi and {Altavilla}, Giuseppe and {Althaus}, Christian and {Alvarez Trujillo}, Luis Alonso and {Amarsi}, Anish and {Ammler-von Eiff}, Matthias and {Am{\^o}res}, Eduardo and {Andrade}, Laerte and {Antoniadis-Karnavas}, Alexandros and {Ant{\'o}nio}, Carlos and {Aparicio del Moral}, Beatriz and {Appolloni}, Matteo and {Arena}, Claudio and {Armstrong}, David and {Aroca Aliaga}, Jose and {Asplund}, Martin and {Audenaert}, Jeroen and {Auricchio}, Natalia and {Avelino}, Pedro and {Baeke}, Ann and {Bailli{\'e}}, Kevin and {Balado}, Ana and {Ballber Balaguer{\'o}}, Pau and {Balestra}, Andrea and {Ball}, Warrick and {Ballans}, Herve and {Ballot}, Jerome and {Barban}, Caroline and {Barbary}, Ga{\"e}le and {Barbieri}, Mauro and {Barcel{\'o} Forteza}, Sebasti{\`a} and {Barker}, Adrian and {Barklem}, Paul and {Barnes}, Sydney and {Barrado Navascues}, David and {Barragan}, Oscar and {Baruteau}, Cl{\'e}ment and {Basu}, Sarbani and {Baudin}, Frederic and {Baumeister}, Philipp and {Bayliss}, Daniel and {Bazot}, Michael and {Beck}, Paul G. and {Belkacem}, Kevin and {Bellinger}, Earl and {Benatti}, Serena and {Benomar}, Othman and {B{\'e}rard}, Diane and {Bergemann}, Maria and {Bergomi}, Maria and {Bernardo}, Pierre and {Biazzo}, Katia and {Bignamini}, Andrea and {Bigot}, Lionel and {Billot}, Nicolas and {Binet}, Martin and {Biondi}, David and {Biondi}, Federico and {Birch}, Aaron C. and {Bitsch}, Bertram and {Bluhm Ceballos}, Paz Victoria and {B{\'o}di}, Attila and {Bogn{\'a}r}, Zs{\'o}fia and {Boisse}, Isabelle and {Bolmont}, Emeline and {Bonanno}, Alfio and {Bonavita}, Mariangela and {Bonfanti}, Andrea and {Bonfils}, Xavier and {Bonito}, Rosaria and {Bonomo}, Aldo Stefano and {B{\"o}rner}, Anko and {Boro Saikia}, Sudeshna and {Borreguero Mart{\'\i}n}, Elisa and {Borsa}, Francesco and {Borsato}, Luca and {Bossini}, Diego and {Bouchy}, Francois and {Bou{\'e}}, Gwena{\"e}l and {Boufleur}, Rodrigo and {Boumier}, Patrick and {Bourrier}, Vincent and {Bowman}, Dominic M. and {Bozzo}, Enrico and {Bradley}, Louisa and {Bray}, John and {Bressan}, Alessandro and {Breton}, Sylvain and {Brienza}, Daniele and {Brito}, Ana and {Brogi}, Matteo and {Brown}, Beverly and {Brown}, David J.~A. and {Brun}, Allan Sacha and {Bruno}, Giovanni and {Bruns}, Michael and {Buchhave}, Lars A. and {Bugnet}, Lisa and {Buldgen}, Ga{\"e}l and {Burgess}, Patrick and {Busatta}, Andrea and {Busso}, Giorgia and {Buzasi}, Derek and {Caballero}, Jos{\'e} A. and {Cabral}, Alexandre and {Cabrero Gomez}, Juan-Francisco and {Calderone}, Flavia and {Cameron}, Robert and {Cameron}, Andrew and {Campante}, Tiago and {Campos Gestal}, N{\'e}stor and {Canto Martins}, Bruno Leonardo and {Cara}, Christophe and {Carone}, Ludmila and {Carrasco}, Josep Manel and {Casagrande}, Luca and {Casewell}, Sarah L. and {Cassisi}, Santi and {Castellani}, Marco and {Castro}, Matthieu and {Catala}, Claude and {Catal{\'a}n Fern{\'a}ndez}, Irene and {Catelan}, M{\'a}rcio and {Cegla}, Heather and {Cerruti}, Chiara and {Cessa}, Virginie and {Chadid}, Merieme and {Chaplin}, William and {Charpinet}, Stephane and {Chiappini}, Cristina and {Chiarucci}, Simone and {Chiavassa}, Andrea and {Chinellato}, Simonetta and {Chirulli}, Giovanni and {Christensen-Dalsgaard}, J{\o}rgen and {Church}, Ross and {Claret}, Antonio and {Clarke}, Cathie and {Claudi}, Riccardo and {Clermont}, Lionel and {Coelho}, Hugo and {Coelho}, Joao and {Cogato}, Fabrizio and {Colom{\'e}}, Josep and {Condamin}, Mathieu and {Conde Garc{\'\i}a}, Fernando and {Conseil}, Simon},
        title = "{The PLATO mission}",
      journal = {Experimental Astronomy},
         year = 2025,
        month = jun,
       volume = {59},
       number = {3},
          eid = {26},
        pages = {26},
          doi = {10.1007/s10686-025-09985-9},
archivePrefix = {arXiv},
       eprint = {2406.05447},
 primaryClass = {astro-ph.IM},
       adsurl = {https://ui.adsabs.harvard.edu/abs/2025ExA....59...26R}
}

@ARTICLE{gaia,
       author = {{Gaia Collaboration} and {Vallenari}, A. and {Brown}, A.~G.~A. and {Prusti}, T. and {de Bruijne}, J.~H.~J. and {Arenou}, F. and {Babusiaux}, C. and {Biermann}, M. and {Creevey}, O.~L. and {Ducourant}, C. and {Evans}, D.~W. and {Eyer}, L. and {Guerra}, R. and {Hutton}, A. and {Jordi}, C. and {Klioner}, S.~A. and {Lammers}, U.~L. and {Lindegren}, L. and {Luri}, X. and {Mignard}, F. and {Panem}, C. and {Pourbaix}, D. and {Randich}, S. and {Sartoretti}, P. and {Soubiran}, C. and {Tanga}, P. and {Walton}, N.~A. and {Bailer-Jones}, C.~A.~L. and {Bastian}, U. and {Drimmel}, R. and {Jansen}, F. and {Katz}, D. and {Lattanzi}, M.~G. and {van Leeuwen}, F. and {Bakker}, J. and {Cacciari}, C. and {Casta{\~n}eda}, J. and {De Angeli}, F. and {Fabricius}, C. and {Fouesneau}, M. and {Fr{\'e}mat}, Y. and {Galluccio}, L. and {Guerrier}, A. and {Heiter}, U. and {Masana}, E. and {Messineo}, R. and {Mowlavi}, N. and {Nicolas}, C. and {Nienartowicz}, K. and {Pailler}, F. and {Panuzzo}, P. and {Riclet}, F. and {Roux}, W. and {Seabroke}, G.~M. and {Sordo}, R. and {Th{\'e}venin}, F. and {Gracia-Abril}, G. and {Portell}, J. and {Teyssier}, D. and {Altmann}, M. and {Andrae}, R. and {Audard}, M. and {Bellas-Velidis}, I. and {Benson}, K. and {Berthier}, J. and {Blomme}, R. and {Burgess}, P.~W. and {Busonero}, D. and {Busso}, G. and {C{\'a}novas}, H. and {Carry}, B. and {Cellino}, A. and {Cheek}, N. and {Clementini}, G. and {Damerdji}, Y. and {Davidson}, M. and {de Teodoro}, P. and {Nu{\~n}ez Campos}, M. and {Delchambre}, L. and {Dell'Oro}, A. and {Esquej}, P. and {Fern{\'a}ndez-Hern{\'a}ndez}, J. and {Fraile}, E. and {Garabato}, D. and {Garc{\'\i}a-Lario}, P. and {Gosset}, E. and {Haigron}, R. and {Halbwachs}, J.-L. and {Hambly}, N.~C. and {Harrison}, D.~L. and {Hern{\'a}ndez}, J. and {Hestroffer}, D. and {Hodgkin}, S.~T. and {Holl}, B. and {Jan{\ss}en}, K. and {Jevardat de Fombelle}, G. and {Jordan}, S. and {Krone-Martins}, A. and {Lanzafame}, A.~C. and {L{\"o}ffler}, W. and {Marchal}, O. and {Marrese}, P.~M. and {Moitinho}, A. and {Muinonen}, K. and {Osborne}, P. and {Pancino}, E. and {Pauwels}, T. and {Recio-Blanco}, A. and {Reyl{\'e}}, C. and {Riello}, M. and {Rimoldini}, L. and {Roegiers}, T. and {Rybizki}, J. and {Sarro}, L.~M. and {Siopis}, C. and {Smith}, M. and {Sozzetti}, A. and {Utrilla}, E. and {van Leeuwen}, M. and {Abbas}, U. and {{\'A}brah{\'a}m}, P. and {Abreu Aramburu}, A. and {Aerts}, C. and {Aguado}, J.~J. and {Ajaj}, M. and {Aldea-Montero}, F. and {Altavilla}, G. and {{\'A}lvarez}, M.~A. and {Alves}, J. and {Anders}, F. and {Anderson}, R.~I. and {Anglada Varela}, E. and {Antoja}, T. and {Baines}, D. and {Baker}, S.~G. and {Balaguer-N{\'u}{\~n}ez}, L. and {Balbinot}, E. and {Balog}, Z. and {Barache}, C. and {Barbato}, D. and {Barros}, M. and {Barstow}, M.~A. and {Bartolom{\'e}}, S. and {Bassilana}, J.-L. and {Bauchet}, N. and {Becciani}, U. and {Bellazzini}, M. and {Berihuete}, A. and {Bernet}, M. and {Bertone}, S. and {Bianchi}, L. and {Binnenfeld}, A. and {Blanco-Cuaresma}, S. and {Blazere}, A. and {Boch}, T. and {Bombrun}, A. and {Bossini}, D. and {Bouquillon}, S. and {Bragaglia}, A. and {Bramante}, L. and {Breedt}, E. and {Bressan}, A. and {Brouillet}, N. and {Brugaletta}, E. and {Bucciarelli}, B. and {Burlacu}, A. and {Butkevich}, A.~G. and {Buzzi}, R. and {Caffau}, E. and {Cancelliere}, R. and {Cantat-Gaudin}, T. and {Carballo}, R. and {Carlucci}, T. and {Carnerero}, M.~I. and {Carrasco}, J.~M. and {Casamiquela}, L. and {Castellani}, M. and {Castro-Ginard}, A. and {Chaoul}, L. and {Charlot}, P. and {Chemin}, L. and {Chiaramida}, V. and {Chiavassa}, A. and {Chornay}, N. and {Comoretto}, G. and {Contursi}, G. and {Cooper}, W.~J. and {Cornez}, T. and {Cowell}, S. and {Crifo}, F. and {Cropper}, M. and {Crosta}, M. and {Crowley}, C. and {Dafonte}, C. and {Dapergolas}, A. and {David}, M. and {David}, P. and {de Laverny}, P. and {De Luise}, F. and {De March}, R.},
        title = "{Gaia Data Release 3. Summary of the content and survey properties}",
      journal = {\aap},
         year = 2023,
        month = jun,
       volume = {674},
          eid = {A1},
        pages = {A1},
          doi = {10.1051/0004-6361/202243940},
archivePrefix = {arXiv},
       eprint = {2208.00211},
 primaryClass = {astro-ph.GA},
       adsurl = {https://ui.adsabs.harvard.edu/abs/2023A&A...674A...1G}
}

@ARTICLE{gravity,
       author = {{GRAVITY Collaboration} and {Abuter}, R. and {Accardo}, M. and {Amorim}, A. and {Anugu}, N. and {{\'A}vila}, G. and {Azouaoui}, N. and {Benisty}, M. and {Berger}, J.~P. and {Blind}, N. and {Bonnet}, H. and {Bourget}, P. and {Brandner}, W. and {Brast}, R. and {Buron}, A. and {Burtscher}, L. and {Cassaing}, F. and {Chapron}, F. and {Choquet}, {\'E}. and {Cl{\'e}net}, Y. and {Collin}, C. and {Coud{\'e} Du Foresto}, V. and {de Wit}, W. and {de Zeeuw}, P.~T. and {Deen}, C. and {Delplancke-Str{\"o}bele}, F. and {Dembet}, R. and {Derie}, F. and {Dexter}, J. and {Duvert}, G. and {Ebert}, M. and {Eckart}, A. and {Eisenhauer}, F. and {Esselborn}, M. and {F{\'e}dou}, P. and {Finger}, G. and {Garcia}, P. and {Garcia Dabo}, C.~E. and {Garcia Lopez}, R. and {Gendron}, E. and {Genzel}, R. and {Gillessen}, S. and {Gonte}, F. and {Gordo}, P. and {Grould}, M. and {Gr{\"o}zinger}, U. and {Guieu}, S. and {Haguenauer}, P. and {Hans}, O. and {Haubois}, X. and {Haug}, M. and {Haussmann}, F. and {Henning}, Th. and {Hippler}, S. and {Horrobin}, M. and {Huber}, A. and {Hubert}, Z. and {Hubin}, N. and {Hummel}, C.~A. and {Jakob}, G. and {Janssen}, A. and {Jochum}, L. and {Jocou}, L. and {Kaufer}, A. and {Kellner}, S. and {Kendrew}, S. and {Kern}, L. and {Kervella}, P. and {Kiekebusch}, M. and {Klein}, R. and {Kok}, Y. and {Kolb}, J. and {Kulas}, M. and {Lacour}, S. and {Lapeyr{\`e}re}, V. and {Lazareff}, B. and {Le Bouquin}, J. -B. and {L{\`e}na}, P. and {Lenzen}, R. and {L{\'e}v{\^e}que}, S. and {Lippa}, M. and {Magnard}, Y. and {Mehrgan}, L. and {Mellein}, M. and {M{\'e}rand}, A. and {Moreno-Ventas}, J. and {Moulin}, T. and {M{\"u}ller}, E. and {M{\"u}ller}, F. and {Neumann}, U. and {Oberti}, S. and {Ott}, T. and {Pallanca}, L. and {Panduro}, J. and {Pasquini}, L. and {Paumard}, T. and {Percheron}, I. and {Perraut}, K. and {Perrin}, G. and {Pfl{\"u}ger}, A. and {Pfuhl}, O. and {Phan Duc}, T. and {Plewa}, P.~M. and {Popovic}, D. and {Rabien}, S. and {Ram{\'\i}rez}, A. and {Ramos}, J. and {Rau}, C. and {Riquelme}, M. and {Rohloff}, R. -R. and {Rousset}, G. and {Sanchez-Bermudez}, J. and {Scheithauer}, S. and {Sch{\"o}ller}, M. and {Schuhler}, N. and {Spyromilio}, J. and {Straubmeier}, C. and {Sturm}, E. and {Suarez}, M. and {Tristram}, K.~R.~W. and {Ventura}, N. and {Vincent}, F. and {Waisberg}, I. and {Wank}, I. and {Weber}, J. and {Wieprecht}, E. and {Wiest}, M. and {Wiezorrek}, E. and {Wittkowski}, M. and {Woillez}, J. and {Wolff}, B. and {Yazici}, S. and {Ziegler}, D. and {Zins}, G.},
        title = "{First light for GRAVITY: Phase referencing optical interferometry for the Very Large Telescope Interferometer}",
      journal = {\aap},
         year = 2017,
        month = jun,
       volume = {602},
          eid = {A94},
        pages = {A94},
          doi = {10.1051/0004-6361/201730838},
archivePrefix = {arXiv},
       eprint = {1705.02345},
 primaryClass = {astro-ph.IM},
       adsurl = {https://ui.adsabs.harvard.edu/abs/2017A&A...602A..94G}
}

@INPROCEEDINGS{pavo,
   author = {{Ireland}, M.~J. and {M{\'e}rand}, A. and {ten Brummelaar}, T.~A. and
	{Tuthill}, P.~G. and {Schaefer}, G.~H. and {Turner}, N.~H. and
	{Sturmann}, J. and {Sturmann}, L. and {McAlister}, H.~A.},
    title = "{Sensitive visible interferometry with PAVO}",
booktitle = {Society of Photo-Optical Instrumentation Engineers (SPIE) Conference Series},
     year = 2008,
   series = {Presented at the Society of Photo-Optical Instrumentation Engineers (SPIE) Conference},
   volume = 7013,
    month = jul,
      doi = {10.1117/12.788386},
   adsurl = {http://cdsads.u-strasbg.fr/abs/2008SPIE.7013E..63I}
}

@ARTICLE{pavo2,
       author = {{White}, T.~R. and {Huber}, D. and {Mann}, A.~W. and {Casagrande}, L. and {Grunblatt}, S.~K. and {Justesen}, A.~B. and {Silva Aguirre}, V. and {Bedding}, T.~R. and {Ireland}, M.~J. and {Schaefer}, G.~H. and {Tuthill}, P.~G.},
        title = "{Interferometric diameters of five evolved intermediate-mass planet-hosting stars measured with PAVO at the CHARA Array}",
      journal = {\mnras},
         year = 2018,
        month = jul,
       volume = {477},
       number = {4},
        pages = {4403-4413},
          doi = {10.1093/mnras/sty898},
archivePrefix = {arXiv},
       eprint = {1804.05976},
 primaryClass = {astro-ph.SR},
       adsurl = {https://ui.adsabs.harvard.edu/abs/2018MNRAS.477.4403W}
}

@ARTICLE{oifits2,
       author = {{Duvert}, Gilles and {Young}, John and {Hummel}, Christian A.},
        title = "{OIFITS 2: the 2nd version of the data exchange standard for optical interferometry}",
      journal = {\aap},
         year = 2017,
        month = jan,
       volume = {597},
          eid = {A8},
        pages = {A8},
          doi = {10.1051/0004-6361/201526405},
archivePrefix = {arXiv},
       eprint = {1510.04556},
 primaryClass = {astro-ph.IM},
       adsurl = {https://ui.adsabs.harvard.edu/abs/2017A&A...597A...8D}
}

@ARTICLE{lachaume2019,
       author = {{Lachaume}, R{\'e}gis and {Rabus}, Markus and {Jord{\'a}n}, Andr{\'e}s and {Brahm}, Rafael and {Boyajian}, Tabetha and {von Braun}, Kaspar and {Berger}, Jean-Philippe},
        title = "{Towards reliable uncertainties in IR interferometry: the bootstrap for correlated statistical and systematic errors}",
      journal = {\mnras},
         year = 2019,
        month = apr,
       volume = {484},
       number = {2},
        pages = {2656-2673},
          doi = {10.1093/mnras/stz114},
archivePrefix = {arXiv},
       eprint = {1901.02879},
 primaryClass = {astro-ph.IM},
       adsurl = {https://ui.adsabs.harvard.edu/abs/2019MNRAS.484.2656L}
}

@misc{jsdc2017,
       author = {{Bourges}, L. and {Mella}, G. and {Lafrasse}, S. and {Duvert}, G. and {Chelli}, A. and {Le Bouquin}, J. -B. and {Delfosse}, X. and {Chesneau}, O.},
        title = "{VizieR Online Data Catalog: JMMC Stellar Diameters Catalogue - JSDC. Version 2 (Bourges+, 2017)}",
 howpublished = {VizieR On-line Data Catalog: II/346.  Originally published in: 2014ASPC..485..223B},
         year = 2017,
        month = jan,
          eid = {II/346},
       adsurl = {https://ui.adsabs.harvard.edu/abs/2017yCat.2346....0B}
}

@INPROCEEDINGS{spicaft,
       author = {{Pannetier}, Cyril and {B{\'e}rio}, Philippe and {Mourard}, Denis and {Rousseau}, Sylvain and {Allouche}, Fatm{\'e} and {Dejonghe}, Julien and {Bailet}, Christophe and {Lecron}, Daniel and {Cassaing}, Fr{\'e}d{\'e}ric and {Le Bouquin}, Jean-Baptiste and {Perraut}, Karine and {Monnier}, John and {Anugu}, Narsireddy and {ten Brummelaar}, Theo},
        title = "{SPICA-FT: the new fringe tracker of the CHARA array}",
    booktitle = {Optical and Infrared Interferometry and Imaging VIII},
         year = 2022,
       editor = {{M{\'e}rand}, Antoine and {Sallum}, Stephanie and {Sanchez-Bermudez}, Joel},
       series = {Society of Photo-Optical Instrumentation Engineers (SPIE) Conference Series},
       volume = {12183},
        month = aug,
          eid = {1218309},
        pages = {1218309},
          doi = {10.1117/12.2628897},
archivePrefix = {arXiv},
       eprint = {2210.09042},
 primaryClass = {astro-ph.IM},
       adsurl = {https://ui.adsabs.harvard.edu/abs/2022SPIE12183E..09P}
}

@INPROCEEDINGS{mroi,
       author = {{Creech-Eakman}, M.~J. and {Romero}, V.~D. and {Haniff}, C.~A. and {Buscher}, D.~F. and {Young}, J.~S. and {Olivares}, A. and {Salcido}, C. and {Altamirano}, J. and {Barrios}, J.~P.~L.~G. and {Collins}, R. and {Cook}, W. and {Farris}, A. and {Fisher}, M. and {Frothingham}, D. and {Gino}, C. and {Giron}, J. and {Giron}, M. and {Haque}, A. and {Hernandez}, J. and {Jorgensen}, A.~M. and {Loskamp}, L.~P. and {Luis}, J.~J.~D. and {Mason}, J.~C. and {Momeni}, H. and {Norouzi}, S. and {Norris}, R. and {Owens}, G. and {Pino}, J. and {Rochelle}, S. and {Santoro}, R. and {Schofield}, I.~S. and {Seneta}, E.~B. and {Tilton}, J. and {Wilson}, D.},
        title = "{Recent progress with the Magdalena Ridge Observatory Interferometer project}",
    booktitle = {Optical and Infrared Interferometry and Imaging IX},
         year = 2024,
       editor = {{Kammerer}, Jens and {Sallum}, Stephanie and {Sanchez-Bermudez}, Joel},
       series = {Society of Photo-Optical Instrumentation Engineers (SPIE) Conference Series},
       volume = {13095},
        month = aug,
          eid = {130950G},
        pages = {130950G},
          doi = {10.1117/12.3021250},
       adsurl = {https://ui.adsabs.harvard.edu/abs/2024SPIE13095E..0GC}
}

@ARTICLE{lebreton2014,
       author = {{Lebreton}, Y. and {Goupil}, M.~J.},
        title = "{Asteroseismology for ``{\`a} la carte'' stellar age-dating and weighing. Age and mass of the CoRoT exoplanet host HD 52265}",
      journal = {\aap},
         year = 2014,
        month = sep,
       volume = {569},
          eid = {A21},
        pages = {A21},
          doi = {10.1051/0004-6361/201423797},
archivePrefix = {arXiv},
       eprint = {1406.0652},
 primaryClass = {astro-ph.SR},
       adsurl = {https://ui.adsabs.harvard.edu/abs/2014A&A...569A..21L}
}

@ARTICLE{Gent2022,
       author = {{Gent}, Matthew Raymond and {Bergemann}, Maria and {Serenelli}, Aldo and {Casagrande}, Luca and {Gerber}, Jeffrey M. and {Heiter}, Ulrike and {Kovalev}, Mikhail and {Morel}, Thierry and {Nardetto}, Nicolas and {Adibekyan}, Vardan and {Silva Aguirre}, V{\'\i}ctor and {Asplund}, Martin and {Belkacem}, Kevin and {del Burgo}, Carlos and {Bigot}, Lionel and {Chiavassa}, Andrea and {Rodr{\'\i}guez D{\'\i}az}, Luisa Fernanda and {Goupil}, Marie-Jo and {Gonz{\'a}lez Hern{\'a}ndez}, Jonay I. and {Mourard}, Denis and {Merle}, Thibault and {M{\'e}sz{\'a}ros}, Szabolcs and {Marshall}, Douglas J. and {Ouazzani}, Rhita-Maria and {Plez}, Bertrand and {Reese}, Daniel and {Trampedach}, Regner and {Tsantaki}, Maria},
        title = "{The SAPP pipeline for the determination of stellar abundances and atmospheric parameters of stars in the core program of the PLATO mission}",
      journal = {\aap},
         year = 2022,
        month = feb,
       volume = {658},
          eid = {A147},
        pages = {A147},
          doi = {10.1051/0004-6361/202140863},
archivePrefix = {arXiv},
       eprint = {2111.06666},
 primaryClass = {astro-ph.SR},
       adsurl = {https://ui.adsabs.harvard.edu/abs/2022A&A...658A.147G}
}

@ARTICLE{di_mauro_2022,
       author = {{Di Mauro}, Maria Pia and {Reda}, Raffaele and {Mathur}, Savita and {Garc{\'\i}a}, Rafael A. and {Buzasi}, Derek L. and {Corsaro}, Enrico and {Benomar}, Othman and {Gonz{\'a}lez Cuesta}, Luc{\'\i}a and {Stassun}, Keivan G. and {Benatti}, Serena and {D'Orazi}, Valentina and {Giovannelli}, Luca and {Mesa}, Dino and {Nardetto}, Nicolas},
        title = "{On the Characterization of GJ 504: A Magnetically Active Planet-host Star Observed by the Transiting Exoplanet Survey Satellite (TESS)}",
      journal = {\apj},
         year = 2022,
        month = nov,
       volume = {940},
       number = {1},
          eid = {93},
        pages = {93},
          doi = {10.3847/1538-4357/ac8f44},
archivePrefix = {arXiv},
       eprint = {2209.12752},
 primaryClass = {astro-ph.SR},
       adsurl = {https://ui.adsabs.harvard.edu/abs/2022ApJ...940...93D}
}

@ARTICLE{valle_2024,
       author = {{Valle}, G. and {Dell'Omodarme}, M. and {Prada Moroni}, P.~G. and {Degl'Innocenti}, S.},
        title = "{Testing the asteroseismic estimates of stellar radii with surface brightness-colour relations and Gaia DR3 parallaxes: Red giants and red clump stars}",
      journal = {\aap},
         year = 2024,
        month = oct,
       volume = {690},
          eid = {A327},
        pages = {A327},
          doi = {10.1051/0004-6361/202451473},
archivePrefix = {arXiv},
       eprint = {2409.10050},
 primaryClass = {astro-ph.SR},
       adsurl = {https://ui.adsabs.harvard.edu/abs/2024A&A...690A.327V}
}

@ARTICLE{campante_2019,
       author = {{Campante}, Tiago L. and {Corsaro}, Enrico and {Lund}, Mikkel N. and {Mosser}, Beno{\^\i}t and {Serenelli}, Aldo and {Veras}, Dimitri and {Adibekyan}, Vardan and {Antia}, H.~M. and {Ball}, Warrick and {Basu}, Sarbani and {Bedding}, Timothy R. and {Bossini}, Diego and {Davies}, Guy R. and {Delgado Mena}, Elisa and {Garc{\'\i}a}, Rafael A. and {Handberg}, Rasmus and {Hon}, Marc and {Kane}, Stephen R. and {Kawaler}, Steven D. and {Kuszlewicz}, James S. and {Lucas}, Miles and {Mathur}, Savita and {Nardetto}, Nicolas and {Nielsen}, Martin B. and {Pinsonneault}, Marc H. and {Reffert}, Sabine and {Silva Aguirre}, V{\'\i}ctor and {Stassun}, Keivan G. and {Stello}, Dennis and {Stock}, Stephan and {Vrard}, Mathieu and {Y{\i}ld{\i}z}, Mutlu and {Chaplin}, William J. and {Huber}, Daniel and {Bean}, Jacob L. and {{\c{C}}elik Orhan}, Zeynep and {Cunha}, Margarida S. and {Christensen-Dalsgaard}, J{\o}rgen and {Kjeldsen}, Hans and {Metcalfe}, Travis S. and {Miglio}, Andrea and {Monteiro}, M{\'a}rio J.~P.~F.~G. and {Nsamba}, Benard and {{\"O}rtel}, Sibel and {Pereira}, Filipe and {Sousa}, S{\'e}rgio G. and {Tsantaki}, Maria and {Turnbull}, Margaret C.},
        title = "{TESS Asteroseismology of the Known Red-giant Host Stars HD 212771 and HD 203949}",
      journal = {\apj},
         year = 2019,
        month = nov,
       volume = {885},
       number = {1},
          eid = {31},
        pages = {31},
          doi = {10.3847/1538-4357/ab44a8},
archivePrefix = {arXiv},
       eprint = {1909.05961},
 primaryClass = {astro-ph.SR},
       adsurl = {https://ui.adsabs.harvard.edu/abs/2019ApJ...885...31C}
}

@ARTICLE{pietrzynski_2013,
       author = {{Pietrzy{\'n}ski}, G. and {Graczyk}, D. and {Gieren}, W. and {Thompson}, I.~B. and {Pilecki}, B. and {Udalski}, A. and {Soszy{\'n}ski}, I. and {Koz{\l}owski}, S. and {Konorski}, P. and {Suchomska}, K. and {Bono}, G. and {Moroni}, P.~G. Prada and {Villanova}, S. and {Nardetto}, N. and {Bresolin}, F. and {Kudritzki}, R.~P. and {Storm}, J. and {Gallenne}, A. and {Smolec}, R. and {Minniti}, D. and {Kubiak}, M. and {Szyma{\'n}ski}, M.~K. and {Poleski}, R. and {Wyrzykowski}, {\L}. and {Ulaczyk}, K. and {Pietrukowicz}, P. and {G{\'o}rski}, M. and {Karczmarek}, P.},
        title = "{An eclipsing-binary distance to the Large Magellanic Cloud accurate to two per cent}",
      journal = {\nat},
         year = 2013,
        month = mar,
       volume = {495},
       number = {7439},
        pages = {76-79},
          doi = {10.1038/nature11878},
archivePrefix = {arXiv},
       eprint = {1303.2063},
 primaryClass = {astro-ph.GA},
       adsurl = {https://ui.adsabs.harvard.edu/abs/2013Natur.495...76P}
}

@ARTICLE{pietrzynski_2019,
       author = {{Pietrzy{\'n}ski}, G. and {Graczyk}, D. and {Gallenne}, A. and {Gieren}, W. and {Thompson}, I.~B. and {Pilecki}, B. and {Karczmarek}, P. and {G{\'o}rski}, M. and {Suchomska}, K. and {Taormina}, M. and {Zgirski}, B. and {Wielg{\'o}rski}, P. and {Ko{\l}aczkowski}, Z. and {Konorski}, P. and {Villanova}, S. and {Nardetto}, N. and {Kervella}, P. and {Bresolin}, F. and {Kudritzki}, R.~P. and {Storm}, J. and {Smolec}, R. and {Narloch}, W.},
        title = "{A distance to the Large Magellanic Cloud that is precise to one per cent}",
      journal = {\nat},
         year = 2019,
        month = mar,
       volume = {567},
       number = {7747},
        pages = {200-203},
          doi = {10.1038/s41586-019-0999-4},
archivePrefix = {arXiv},
       eprint = {1903.08096},
 primaryClass = {astro-ph.GA},
       adsurl = {https://ui.adsabs.harvard.edu/abs/2019Natur.567..200P}
}

@ARTICLE{graczyk_2020,
       author = {{Graczyk}, Dariusz and {Pietrzy{\'n}ski}, Grzegorz and {Thompson}, Ian B. and {Gieren}, Wolfgang and {Zgirski}, Bart{\l}omiej and {Villanova}, Sandro and {G{\'o}rski}, Marek and {Wielg{\'o}rski}, Piotr and {Karczmarek}, Paulina and {Narloch}, Weronika and {Pilecki}, Bogumi{\l} and {Taormina}, Monica and {Smolec}, Rados{\l}aw and {Suchomska}, Ksenia and {Gallenne}, Alexandre and {Nardetto}, Nicolas and {Storm}, Jesper and {Kudritzki}, Rolf-Peter and {Ka{\l}uszy{\'n}ski}, Miko{\l}aj and {Pych}, Wojciech},
        title = "{A Distance Determination to the Small Magellanic Cloud with an Accuracy of Better than Two Percent Based on Late-type Eclipsing Binary Stars}",
      journal = {\apj},
         year = 2020,
        month = nov,
       volume = {904},
       number = {1},
          eid = {13},
        pages = {13},
          doi = {10.3847/1538-4357/abbb2b},
archivePrefix = {arXiv},
       eprint = {2010.08754},
 primaryClass = {astro-ph.GA},
       adsurl = {https://ui.adsabs.harvard.edu/abs/2020ApJ...904...13G}
}

@ARTICLE{riess_2022,
       author = {{Riess}, Adam G. and {Yuan}, Wenlong and {Macri}, Lucas M. and {Scolnic}, Dan and {Brout}, Dillon and {Casertano}, Stefano and {Jones}, David O. and {Murakami}, Yukei and {Anand}, Gagandeep S. and {Breuval}, Louise and {Brink}, Thomas G. and {Filippenko}, Alexei V. and {Hoffmann}, Samantha and {Jha}, Saurabh W. and {D'arcy Kenworthy}, W. and {Mackenty}, John and {Stahl}, Benjamin E. and {Zheng}, WeiKang},
        title = "{A Comprehensive Measurement of the Local Value of the Hubble Constant with 1 km s$^{-1}$ Mpc$^{-1}$ Uncertainty from the Hubble Space Telescope and the SH0ES Team}",
      journal = {\apjl},
         year = 2022,
        month = jul,
       volume = {934},
       number = {1},
          eid = {L7},
        pages = {L7},
          doi = {10.3847/2041-8213/ac5c5b},
archivePrefix = {arXiv},
       eprint = {2112.04510},
 primaryClass = {astro-ph.CO},
       adsurl = {https://ui.adsabs.harvard.edu/abs/2022ApJ...934L...7R}
}

@ARTICLE{lundkvist2025,
       author = {{Lundkvist}, M.~S. and {Larsen}, J.~R. and {Li}, Y. and {Winther}, M.~L. and {Bedding}, T.~R. and {Kjeldsen}, H. and {White}, T.~R. and {Nielsen}, M.~B. and {Buldgen}, G. and {Guillaume}, C. and {Stokholm}, A.~L. and {Huber}, D. and {R{\o}rsted}, J.~L. and {Mani}, P. and {Grundahl}, F.},
        title = "{Asteroseismic investigation of HD 140283: The Methuselah star}",
      journal = {\aap},
         year = 2025,
        month = nov,
       volume = {703},
          eid = {A232},
        pages = {A232},
          doi = {10.1051/0004-6361/202556292},
archivePrefix = {arXiv},
       eprint = {2510.11532},
 primaryClass = {astro-ph.SR},
       adsurl = {https://ui.adsabs.harvard.edu/abs/2025A&A...703A.232L}
}

@BOOK{strehl,
       author = {{Born}, Max and {Wolf}, Emil},
        title = "{Principles of Optics}",
         year = 2019,
          doi = {10.1017/9781108769914},
       adsurl = {https://ui.adsabs.harvard.edu/abs/2019prop.book.....B}
}

@INPROCEEDINGS{berger2003,
       author = {{Berger}, David H. and {ten Brummelaar}, Theo A. and {Bagnuolo}, Jr., William G. and {McAlister}, Harold A.},
        title = "{Preliminary results from the longitudinal dispersion compensation system for the CHARA array}",
    booktitle = {Interferometry for Optical Astronomy II},
         year = 2003,
       editor = {{Traub}, Wesley A.},
       series = {Society of Photo-Optical Instrumentation Engineers (SPIE) Conference Series},
       volume = {4838},
        month = feb,
        pages = {974-982},
          doi = {10.1117/12.457161},
       adsurl = {https://ui.adsabs.harvard.edu/abs/2003SPIE.4838..974B}
}

@article{anugu2026,
author = {Narsireddy Anugu and Nils H. Turner and Theo A. ten Brummelaar and Gail H. Schaefer and Philippe B{\'e}rio and Christopher D. Farrington and Becky Flores and Douglas R. Gies and Stefan Kraus and Edgar R. Ligon and Olli W. Majoinen and John D. Monnier and Denis Mourard and Nicholas  J. Scott and Norman  L. Vargas},
title = {{CHARA Array delay lines: upgrades, performance, and future directions}},
volume = {12},
journal = {Journal of Astronomical Telescopes, Instruments, and Systems},
number = {1},
publisher = {SPIE},
pages = {015008},
year = {2026},
doi = {10.1117/1.JATIS.12.1.015008},
URL = {https://doi.org/10.1117/1.JATIS.12.1.015008}
}

@INPROCEEDINGS{anugu2020,
       author = {{Anugu}, Narsireddy and {ten Brummelaar}, Theo and {Turner}, Nils H. and {Anderson}, Matthew D. and {Le Bouquin}, Jean-Baptiste and {Sturmann}, Judit and {Sturmann}, Laszlo and {Farrington}, Chris and {Vargas}, Norm and {Majoinen}, Olli and {Ireland}, Michael J. and {Monnier}, John D. and {Mourard}, Denis and {Schaefer}, Gail and {Gies}, Douglas R. and {Ridgway}, Stephen T. and {Kraus}, Stefan and {Petit}, Cyril and {Tallon}, Michel and {Lim}, Caroline B. and {Berio}, Philippe},
        title = "{CHARA array adaptive optics: complex operational software and performance}",
    booktitle = {Optical and Infrared Interferometry and Imaging VII},
         year = 2020,
       editor = {{Tuthill}, Peter G. and {M{\'e}rand}, Antoine and {Sallum}, Stephanie},
       series = {Society of Photo-Optical Instrumentation Engineers (SPIE) Conference Series},
       volume = {11446},
        month = dec,
          eid = {1144622},
        pages = {1144622},
          doi = {10.1117/12.2561560},
archivePrefix = {arXiv},
       eprint = {2012.11667},
 primaryClass = {astro-ph.IM},
       adsurl = {https://ui.adsabs.harvard.edu/abs/2020SPIE11446E..22A}
}

@ARTICLE{chelli2016,
       author = {{Chelli}, Alain and {Duvert}, Gilles and {Bourg{\`e}s}, Laurent and {Mella}, Guillaume and {Lafrasse}, Sylvain and {Bonneau}, Daniel and {Chesneau}, Olivier},
        title = "{Pseudomagnitudes and differential surface brightness: Application to the apparent diameter of stars}",
      journal = {\aap},
         year = 2016,
        month = may,
       volume = {589},
          eid = {A112},
        pages = {A112},
          doi = {10.1051/0004-6361/201527484},
archivePrefix = {arXiv},
       eprint = {1604.07700},
 primaryClass = {astro-ph.SR},
       adsurl = {https://ui.adsabs.harvard.edu/abs/2016A&A...589A.112C}
}

@ARTICLE{ibanez2026,
       author = {{Ibañez Bustos}, R. and {Mourard}, D. and {Nardetto}, N.},
        title = "{Interferometric Survey of Stellar Parameters - Towards homogeneous FGK stars parameters and surface-brightness color relation in the context of PLATO space mission}",
      journal = {\aap},
         year = 2026,
       volume = {submitted},
}

@ARTICLE{nayeem2026,
       author = {{Ebrahimkutty}, N. and {Mourard}, D. and {Berio}, P.},
        title = "{Interferometric Survey of Stellar Parameters: Limb darkening study
with the "Pipeline for Interferometric Measurements of Stars"
(PIMS)}",
      journal = {\aap},
         year = 2026,
       volume = {submitted},
}

@ARTICLE{juraj2026,
       author = {{Jonák}, J. and {Mourard}, D. and {Monnier}, J.},
        title = "{Interferometric Survey of Stellar Parameters:\\ Mass of the metallic A-type binary $\beta$ Aur}",
      journal = {\aap},
         year = 2026,
       volume = {in press},
}

\begin{appendix}

\section{Variance of the ratio of two independent random variables}
\label{sec:ratio}
The variance of the ratio of two independent random variables $Z=X/Y$ is computed with the following formula:
\begin{itemize}
    \item if $S(Y)>4$, the 4th order approximation of Winzer is applied: \begin{equation} Var(Z)=Z^2(\frac{1}{S(X)^2}+\frac{1}{S(Y)^2}-\frac{1}{S(Y)^4}+\frac{3}{S(X)^2S(Y)^2})\end{equation} where \begin{equation}S(X)=\frac{X}{\sigma_X}\end{equation} and \begin{equation}S(Y)=\frac{Y}{\sigma_Y}\end{equation}
    \item if $S(Y)<2$, the asymmetric "modified-triangular" distribution is applied: \begin{equation}Var(Z)=\frac{(Max(Z)-Min(Z))^2}{4}\end{equation} where \begin{equation}Max(Z)=\frac{X+\sigma_X}{Y-\sigma_Y}\end{equation} and \begin{equation} Min(Z)=\frac{X-\sigma_X}{Y+\sigma_Y}\end{equation}
    \item if $2<S(Y)<4$, a linear combination of both preceding methods is applied.
\end{itemize}

\section{List of acronyms}
\label{sec:acronyms}
Table \ref{tab:acronyms} presents the main technical acronyms used in this paper.

\begin{table*}[]
    \centering
    \small
    \caption{List of acronym used in this paper.}
    \begin{tabular}{p{2.5cm}|p{5cm}|p{8cm}}\hline
    Acronym & Definition & Details\\\hline\hline
    ADC & Atmospheric Dispersion Compensator & Device installed in each SPICA-VIS beam to correct the atmospheric refraction before the injection in the single-mode fibers\\\hline 

    BSP & Beam Splitters & The six beam splitters deviating 10\% of the light of each beam to the IPDET detector for the control of the image and pupil planes\\\hline

    CHARA Array & Center for High Angular Resolution Astronomy & the CHARA Array\\\hline

    COL & Collimating device & Optical device permitting to collimate the beams after the FTT mirror and before the INJ Module\\\hline

    DDL & Differential Delay Lines & The differential delay lines of the SPICA-VIS instrument for the cophasing and the compensation of the chromatic offsets\\\hline

    FBI & Fiber Back Illumination & Alignment laser and its movable device to retro feed the SPICA-VIS single-mode fibers for alignment purpose\\\hline

    FOC & Focalization device & Optical system permitting to form the intermediate image plane in the SPICA-VIS instrument\\\hline

    FTT & Fast Tip Tilt & XY-piezo stage for the fast stabilization of each of the SPICA-VIS image at the entrance of the injection module\\\hline
    
    INJ & Injection module & Optical device permitting the injection of the CHARA beams in the single-mode fibers of the SPICA-VIS instrument\\\hline
    
    IPDET & Image Pupil Detector & Detector of the SPICA-VIS instrument for the control of the image and pupil planes\\\hline
    
    LDC & Longitudinal Dispersion Compensator & Visible and Infrared variable glass thickness for the correction of the longitudinal dispersion due to the difference of air paths\\\hline
    
    OPLE & Optical Path Length Equalizer & The CHARA delay lines.\\\hline

    PDC & Polarization Device Compensation & The optical module permitting to compensate the differential polarization between the six fibers of the SPICA-VIS instrument\\\hline
    
    PERI & Periscope & Device installed on the CHARA visible table and permitting (when DOWN) to send the CHARA beams to SPICA-VIS, or to send (when UP) the CHARA reference beams back to the telescopes for alignment and cophasing\\\hline 

    PIS & Pupil Image System & Switchable optical lens at the entrance of the IPDET detector to image either the pupil or the image plane\\\hline 

    PUP & Pupil device & Optical system permitting to stabilize each of the CHARA pupils in lateral and longitudinal directions\\\hline

    RFL & Retro reflectors & Six corner cubes used to send the FBI light reflected by the BSP to the IPDET detector, for internal alignment of each of the SPICA-VIS beam\\\hline
    
    SCIDET & Science Detector & Detector of the SPICA-VIS instrument\\\hline

    SHU & Shutters & Shutters installed in the SPICA-VIS instrument and permitting to close independently any of the six beams.\\\hline
    
    SPICA & Stellar Parameters and Images with a Cophased Array & the SPICA project\\\hline
    
    SPICA-FT & SPICA Fringe Tracker & The fringe tracker of the SPICA project \\\hline
    
    SPICA-VIS & SPICA Visible Instrument & The visible combiner of the SPICA project \\\hline
    
    STS & Six Telescope Simulator & A CHARA Array subsystem that simulates the light from all six telescopes for alignment, cophasing, calibration, and testing of interferometric instruments\\\hline
    
    STST & Six Telescope Star Tracker & A CHARA Array subsystem that allows a slow guiding in the infrared for the stabilization of the alignment in the instruments\\\hline
    
    VIS-FOLD & Visible Fold Optics & Movable device permitting to switch from the telescope beams to the STS beams, for the visible instrument\\\hline
    
    VLDC & Visible Longitudinal Dispersion Compensator & Additional variable glass thickness for an optimal correction of the longitudinal dispersion in the visible\\\hline

    \end{tabular}
    \label{tab:acronyms}
\end{table*}

\end{appendix}

\end{document}